\documentclass{aa}
\usepackage{colortbl}
\usepackage{graphicx}
\usepackage{verbatim}

\usepackage[colorlinks=true,linkcolor=blue,citecolor=blue]{hyperref}

\usepackage{txfonts}
\usepackage{orcidlink}

\usepackage[noabbrev, nameinlink]{cleveref} 
\hypersetup{
    colorlinks=true,
    linkcolor=blue,
    filecolor=magenta,      
    urlcolor=blue,
    pdftitle={Overleaf Example},
    pdfpagemode=FullScreen,
    }

\Crefname{section}{Sect.}{Sects.}
\Crefname{figure}{Fig.}{Figs.}
\Crefname{equation}{Eq.}{Eqs.}

\let\em\it
\usepackage{amstext}

\begin{document} 

\title{Hot exozodiacal dust around Fomalhaut:\\ The MATISSE perspective\thanks{Based on observations collected at the European Southern Observatory under the ESO program 109.23HL.001}}
        
\author{K. Ollmann\inst{\ref{inst1}}\ \orcidlink{0009-0003-6954-5252}
\and F. Kirchschlager\inst{\ref{inst2}}\ \orcidlink{0000-0001-9511-3371}
\and T. A. Stuber\inst{\ref{inst3}}\ \orcidlink{0000-0003-2185-0525}
\and K. Tsishchankava\inst{\ref{inst1}}\ \orcidlink{0009-0002-9371-0740}
\and  A. Matter\inst{\ref{inst4}}\
\and S. Ertel\inst{\ref{inst3}, \ref{inst5} }\ \orcidlink{0000-0002-2314-7289}
\and T. D. Pearce\inst{\ref{inst7}\ \orcidlink{0000-0001-5653-5635}}
\and A. V. Krivov\inst{\ref{inst6}}\
\and S. Wolf\inst{\ref{inst1}}\ \orcidlink{0000-0001-7841-3452}}

\institute{\label{inst1}Institut für Theoretische Physik und Astrophysik, Christian-Albrechts-Universität zu Kiel, Leibnizstr. 15, 24118 Kiel, Germany\\
\email{kollmann@astrophysik.uni-kiel.de}
\and {\label{inst2}Sterrenkundig Observatorium, Ghent University, Krĳgslaan 281-S9, B-9000 Gent, Belgium}
\and {\label{inst3}Department of Astronomy and Steward Observatory, The University of Arizona, 933 North Cherry Ave, Tucson, AZ 85721, USA}
\and {\label{inst4}Laboratoire Lagrange, Observatoire de la C\^{o}te d’Azur, CNRS, Universit\'{e} C\^{o}te d’Azur, Boulevard de l’Observatoire, CS 34229, F-06304 Nice Cedex 4, France}
\and {\label{inst5}Large Binocular Telescope Observatory, The University of Arizona, 933 North Cherry Ave, Tucson, AZ 85721, USA} 
\and {\label{inst6}Astrophysikalisches Institut und Universitätssternwarte, Friedrich-Schiller-Universität Jena, Schillergässchen 2–3, 07745 Jena, Germany} 
\and {\label{inst7}Department of Physics, University of Warwick, Gibbet Hill Road, Coventry CV4 7AL, UK}}

    \abstract
{Excess over the stellar photospheric emission of main-sequence stars has been found in interferometric near-infrared observations, and is attributed to the presence of hot exozodiacal dust (HEZD). As part of our effort to detect and characterize HEZD around the nearby A3 V star Fomalhaut, we carried out the first interferometric observations with the MATISSE instrument at the VLTI in the photometric bands \(L\) and \(M\) for the Fomalhaut system.}
{We investigate whether the new MATISSE data indicate the presence of HEZD around Fomalhaut. If detected, we aim to constrain the  dust grain size, location, dust species, and total dust mass  based on these data. We also investigate if the possibly detected circumstellar radiation could have an alternative explanation.}
{Assuming a dust distribution either as a narrow ring or spherical shell for modeling the HEZD, we aim to constrain the HEZD parameters by generating visibilities and fitting them to the MATISSE data using different approaches.}  
{The MATISSE  \(L\) band data provide a marginal detection of  circumstellar radiation, potentially caused by the presence of HEZD, which is only the second detection of HEZD emission in the \(L\) band. An analysis of the data with different fitting approaches showed that the best-fit values for the HEZD parameters are consistent with those of previous Fomalhaut observations, which again underlines the functionality of MATISSE. We derived the following best-fit HEZD parameter values:  Assuming a dust ring, it would have an inner ring radius of \(0.11\ \mathrm{au}\), an outer ring radius of \(0.12\ \mathrm{au}\), a narrow dust grain size distribution around a dust grain radius constrained by \(0.53\ \mu\mathrm{m}\), and a total dust mass of  \(3.25\times 10^{-10}\ \rm{M}_{\oplus}\). However, even with an additional consideration of previous VINCI (\(K\) band) and KIN (\(N\) band) measurements, we cannot further tighten the constraints for the HEZD properties than in previous Fomalhaut studies. Because our data cannot directly constrain the morphology of the excess radiation source, even the presence of a stellar companion can reproduce the detected marginal visibility deficit. Moreover, the MATISSE data neither imply nor exclude the existence of a double ring structure close to Fomalhaut. Finally, the results indicate that the choice of the geometric model has a more significant impact on the derived dust-to-star flux ratio than the specific fitting approach applied.}
{Since different dust-to-star flux ratios can result from the applied fitting approaches, this also has an impact on the parameter values of the HEZD around Fomalhaut. This circumstance should also be investigated for HEZD systems analyzed so far only with the fitting approach usually applied. Moreover, further NIR and MIR data  are required for a more comprehensive description of the emission originating in the close vicinity of Fomalhaut.}
        
\keywords{circumstellar matter- zodiacal dust- techniques: interferometric- infrared: planetary systems- interplanetary medium
}
\titlerunning{HEZD around Fomalhaut: The MATISSE perspective}
\authorrunning{Ollmann et al.}
\maketitle

\section{Introduction}\label{Intro}
Based on a visibility deficit compared to the stellar photosphere, observations of about two dozen main-sequence stars with high precision and high angular resolution (\(\approx 0.01\) as) long-baseline interferometry have revealed a strong near-infrared (NIR) excess (e.g., \citealt{Absil2006, Absil2013}; \citealt{Ertel}; \citealt{Nunez}) at a percent level with respect to the photospheric emission. This excess has been attributed to the presence of hot exozodiacal dust (HEZD) located in the vicinity of their host stars (spectral types G to A) at or close to the sublimation radius (\citealt{Di Folco}; \citealt{Defrere}). Although these excesses are attributed to submicron-sized dust grains, larger dust particles might have an additional impact on the observational appearance of HEZD (\citealt{Stuber}). The corresponding mid-infrared (MIR) excess in the \(N\) band is much weaker and often not significant (\citealt{Millan-Gabet}; \citealt{Mennesson}). It was found that carbonaceous grains are consistent with the observations. In contrast, the contribution of silicate grains must be negligible since the presence of the prominent silicate feature in the MIR would otherwise contradict the observed weak MIR excess (\citealt{Absil2006}; \citealt{Akeson}; \citealt{Kirchschlager2017}). One strong motivation to study HEZD is that its presence may also offer a way to constrain the architecture and dynamics of the planetary system and that HEZD has the potential to harm future attempts to image and characterize Earth-like exoplanets (e.g., \citealt{Kral}). In particular, the presence of HEZD could also interfere with the polarimetric characterization of close-in exoplanets (\citealt{Ollmann}).\\
\\ 
The origin and physics of HEZD around main-sequence stars is not yet understood because such submicron-sized grains should either rapidly sublimate or be blown out by radiation pressure (e.g., \citealt{Backman}). Therefore, either HEZD has to be replenished on relatively short timescales by exocometary or planetesimal collisions (e.g., \citealt{Bonsor2012,  Bonsor2014}; \citealt{Faramaz}; \citealt{Sezestre}; \citealt{Rigley}; \citealt{Pearce}), or it has to be trapped in the stellar vicinity by gas or magnetic fields (e.g., \citealt{Lebreton}; \citealt{Su2013}; \citealt{Rieke}; \citealt{Stamm}; \citealt{Kimura}; \citealt{Pearce2020}). These and other mechanisms have been proposed, but their relative importance for different systems (e.g., stellar spectral type, amount of cold dust further out in the system, planetary system architecture and dynamics), as well as their combined capacity for explaining the phenomenon of HEZD remains debated (see \citealt{Pearce}).\\
\\
The A3 V star Fomalhaut located at a distance of  7.704 \(\pm\) 0.004 pc (\citealt{Reyle}) with an effective temperature of 8650 K (\citealt{Mamajek}) is one of the nearest systems with a strong NIR excess, allowing excellent spatial resolution (\citealt{MacGregor}) of 13 mas that corresponds to 0.1 au. Observations in the far-IR and submillimeter range have shown a belt of cold dust concentrated at around 140 AU from the star with an eccentric shape, which may or may not be related to the nearby object Fomalhaut b (e.g., \citealt{Kalas2005}; \citealt{Absil}; \citealt{Rieke}; \citealt{Pearce2021}). The first long-baseline interferometric observations of Fomalhaut in the NIR (\citealt{DiFolco2004})  were performed in the \(K\) band with the Very Large Telescope Interferometer (VLTI;  \citealt{Schoeller}; \citealt{Schoeller2003}) using the VLT Interferometer Commissioning Instrument (VINCI; \citealt{Kervella2000, Kervella2003}) and were followed by studies that constrained the corresponding HEZD properties (e.g., \citealt{Absil}; \citealt{Lebreton}). A separate analysis of observations with the Keck Interferometer Nuller (KIN; \citealt{Serabyn}; \citealt{Colavita}) suggested the presence of a small excess between 8 and 11 \(\mu\)m (\citealt{Mennesson}) that is likely the short-wavelength tail of the unresolved cooler excess detected at slightly longer MIR wavelengths (\citealt{Stapelfeldt}; \citealt{Su2013}). Combined spectrophotometric measurements with the space telescopes Herschel and Spitzer (\citealt{Acke}) have given rise to one possible interpretation that this HEZD may be produced by the release of small carbon grains following the disruption of dust aggregates originating from the warm component of the Fomalhaut debris disk. Recent James Webb Space Telescope (JWST; \citealt{Gardner}) observations have also resolved an intermediate belt located around 100 au and an inner belt closer to the star (\citealt{Gaspar}).\\
\\
In this article, we present the first interferometric observation of Fomalhaut in the \(LM\) band. This observation was performed with the Multi AperTure mid-Infrared SpectroScopic Experiment (MATISSE; \citealt{Lopez, Lopez2022}), which is a second-generation instrument at the VLTI that operates in the photometric bands \(L, M,\) and \(N\). Fomalhaut would be only the second object for which around HEZD radiation in the \(L\) band would have been detected, after the F6 IV-V star \(\kappa\) Tuc  (\citealt{Kirchschlager2020}). By extending the observing wavelength range to the \(LM\) band, we aim to more tightly constrain the particle sizes, the total dust mass, the dust species, the location of the HEZD in the Fomalhaut system, and the interconnection between hot and warm dust in relation to previous Fomalhaut studies.\\
\\
This article is organized as follows: We provide an overview of the measurement and the data reduction process for the \(LM\) band \mbox{MATISSE} data in \Cref{matissesection}. Subsequently, we outline two of the applied HEZD fitting approaches (one-step approach and the usual two-step approach) and define a HEZD model in \Cref{model} aimed at constraining the possible parameter space for HEZD around Fomalhaut based on the MATISSE data  (\Cref{results}). We discuss our results in \Cref{discussion}. Our findings are summarized in \Cref{summary}.

\section{MATISSE data}\label{matissesection}
\begin{figure}[t!]
\includegraphics[scale=0.29]{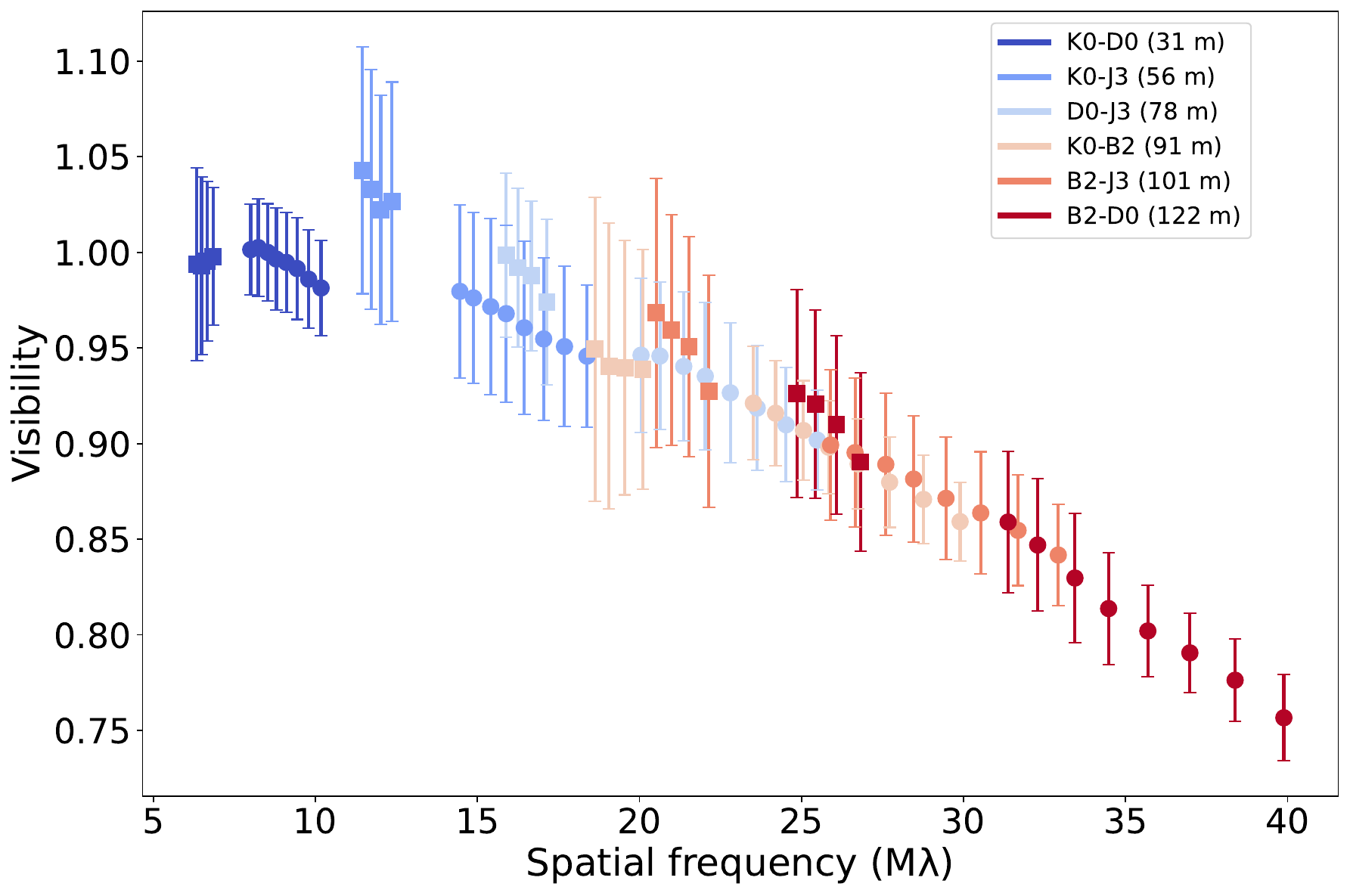}
\caption{Calibrated and merged measurement results  at the different baselines with dependence on the spatial frequency (Mega-\(\lambda\)). The dots denote the \(L\) band data and the squares the \(M\) band data, respectively (see \Cref{matissesection} for details).}\label{Fig first}
\end{figure}

Fomalhaut (HD 216956) was observed on 26 August 2022 as part of a mini-survey of hot exozodiacal disks (ESO programs 109.23HL.001, 0109.C-0706(A); PI: F. Kirchschlager). A CAL-SCI-CAL sequence was carried out (HD 217484 and HD 219784 as calibrators) with MATISSE in medium spectral resolution (\(R=506\)), and the medium AT configuration (projected baselines ranging from \(\sim 31\) m to \(\sim 122\) m) was used for an allocated time of 1.5 h and a seeing of 0.7". Moreover, MATISSE was used in GRA4MAT mode (\citealt{Woillez2019, Woillez2024}), where the GRAVITY fringe tracker (\citealt{Gravity}; \citealt{Lacour}) was used to stabilize fringes for MATISSE.\\
\\
The MATISSE observation sequence is composed of twelve one-minute exposures in which interferometric and photometric data are acquired simultaneously. The first series of four exposures is taken without chopping in the \(LM\) band, while the following eight exposures in the \(LMN\) band are executed with chopping between the target and the sky.
Each exposure is taken in one of four configurations of two beam commuting devices (BCDs), which switch the beams of the telescopes pairwise. This process helps in removing instrumental systematic effects on the phase when the exposures are merged.\\
We used the MATISSE pipeline version 1.7.5 to reduce the raw data. Given the scope of our study and to improve the signal-to-noise ratio (S/N) of our data, we smoothed them down to a low spectral resolution (\(R=30\)) by applying a sliding average on the fringes. The resulting OIFITS (\citealt{Duvert}) files (version 2) contain uncalibrated interferometric observables,
including six dispersed (squared) visibilities and four closure phase measurements per exposure, with 1560 spectral elements within the range of \(2.76 - 4.96\ \mu\mathrm{m}\), knowing that each true spectral channel is sampled by about five pixels on the \(LM\) band detector. The measurements were then calibrated using our two calibrators, whose diameters (about 1.95 mas for HD 217484 and about 2.31 mas for HD 219784) were taken from the JSDC version 2 (\citealt{Chelli}). In particular, the visibility calibration was performed by dividing the raw squared visibilities of Fomalhaut by the mean of the calibrators squared visibilities, corrected for their diameter. For our study, we only considered the \(L\) and \(M\) band data since Fomalhaut is too faint for absolute visibility measurements in the \(N\) band with the ATs. We further restricted the data to the wavelength range of \( \sim 3.0 - 4.0\ \mu\mathrm{m}\) for the \(L\) band and of \(\sim 4.5 - 5.0\ \mu\mathrm{m}\) for the \(M\) band to exclude the band edges, which are affected by telluric absorption, atmospheric dispersion, and the non-transmissive spectral window in the range of \(\sim 4.2 - 4.5\ \mu\mathrm{m}\). With such a filtered range, we ensured good coverage of the \(LM\) bands while only keeping higher quality data (S/N \(> 10\) per spectral channel).\\  Moreover, we only considered the chopped \(LM\) band data, which benefit from a better thermal background subtraction and thus provide a more accurate photometry estimation until for wavelengths up to \(5.0\ \mu\mathrm{m}\).
 The errors of the data points contain the short-term error affecting the individual one-minute exposures of each target (CAL and SCI), the propagated error on the calibrator diameter, as computed by the pipeline, and a longer-term error represented by the standard deviation between the merged exposures. Since  only one CAL-SCI-CAL sequence was performed, the broadband calibration error associated with the stability of the \(L\) and \(M\) bands transfer function over the night was taken to be 2\%, as estimated in good seeing conditions from the MATISSE commissioning (\citealt{Lopez2022}). However, this broadband calibration error appeared to be lower than the data errors per spectral channel and was thus neglected in our error budget. The errors are mostly systematic uncertainties because the data points show little scatter; they follow each other closely for each baseline with minimal noise. However, the uncertainties are much larger than this minimal noise. \\
\\
For our subsequent modeling work, a binning of the measurements and corresponding errors was applied to reconstruct the true spectral channels in low resolution. This resulted in eight calibrated visibility and closure phase measurements between \(\sim\) 3 and 4  \(\mu\mathrm{m}\), and four measurements between \(\sim\) 4.5 and 5 \(\mu\mathrm{m}\). The calibrated and binned visibilities, the Fourier \(uv\) coverage (\(u\) and \(v\) as the spatial frequencies) and the calibrated and binned closure phases are shown in \Cref{Fig first,coverage,Fig 3}. Considering \Cref{Fig first} and \Cref{Fig 3} for the \(M\) band data, the larger error bars and deviation to the \(L\) band data are expected  due to a lower S/N and the stronger contribution of the thermal background, which amplifies the effect of an imperfect chopping.

\begin{figure}[t!]
\includegraphics[scale=0.43]{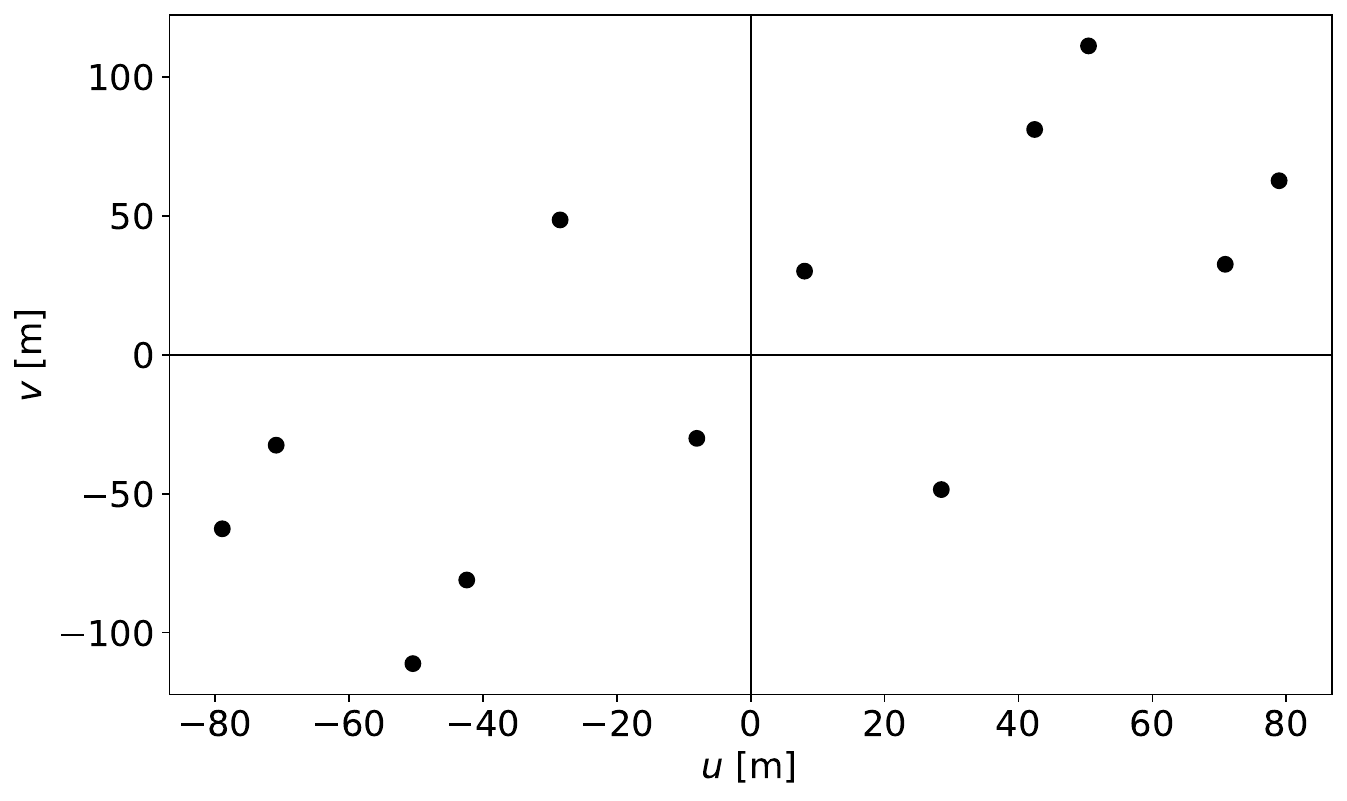}
\caption{Sampling of the Fourier \(uv\) plane for the data set (see \Cref{matissesection} for details).}\label{coverage}
\end{figure}\label{coverage}

\begin{figure}[t!]
\includegraphics[scale=0.22]{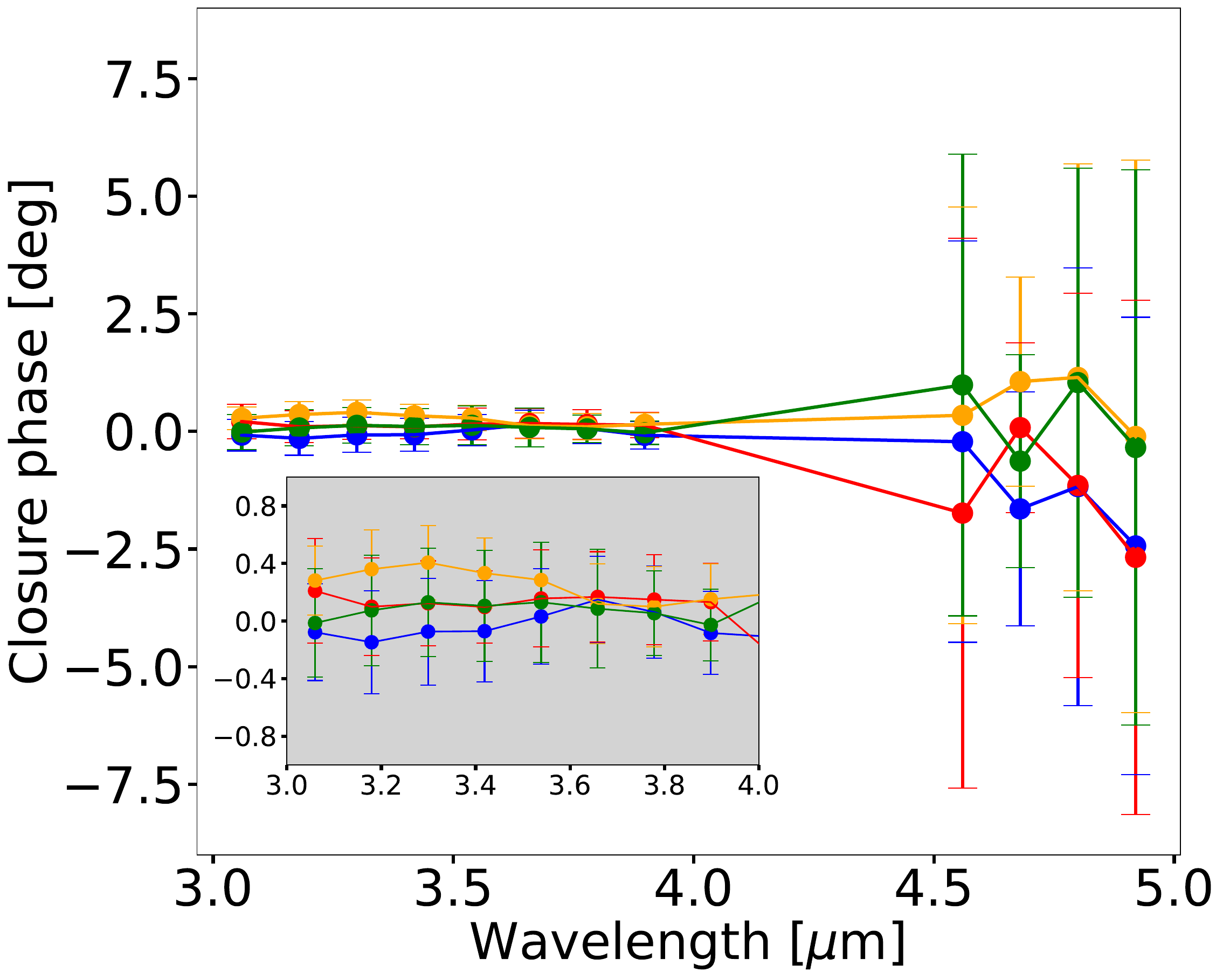}
\caption{Closure phases for the four triangles of the observation (see \Cref{matissesection} and \Cref{analysis} for details).}\label{Fig 3}
\end{figure}

\section{Modeling approach}\label{model}
 We outline two of our applied fitting approaches (see \Cref{bootstrappingmethod} and \Cref{Network} for the outlines of the other applied approaches) in \Cref{procedure} and \Cref{procedure2}. Finally, we describe the HEZD model that we applied to constrain the properties of the HEZD around Fomalhaut in \Cref{procedure3}.

\subsection{Two-step approach}\label{procedure}
The following two-step fitting approach has been used in previous studies (e.g., \citealt{Absil2006}; \citealt{Di Folco}; \citealt{Absil}; \citealt{Defrere}; \citealt{Lebreton}; \citealt{Kirchschlager2020}) to constrain the HEZD model parameter in combination with the \(\chi_{\mathrm{red}}^{2}\) method (see \Cref{procedure3}). In the first step, the visibilities are fit by a model consisting of a limb-darkened stellar photosphere surrounded by a uniformly distributed circumstellar radiation filling the entire field of view (FOV) of the interferometer. The dust-to-star flux ratio \(f_{2}\) is the only free parameter of this model. In the second step, the HEZD parameters are constrained by fitting the significant flux ratios derived in the first step applying the model of an optically thin, geometrically narrow dust ring. Thus, these two steps are geometrically inconsistent. The visibility of the limb-darkened photosphere is given as (\citealt{Hanbury}; \citealt{Kirchschlager2020})
\begin{equation}\label{star} V_{\star}(B,\lambda)=\frac{6}{3-\mu_{\lambda}}\left(\left(1-\mu_{\lambda}\right)\frac{J_{1}(z)}{z}+\mu_{\lambda}\sqrt{\frac{\pi}{2}}\frac{J_{1.5}(z)}{z^{1.5}}\right),\ z=\frac{\pi\Theta_{\star}B}{\lambda},\end{equation}
 with \(\mu_{\lambda}=0.13\) as the linear limb-darkening coefficient for Fomalhaut in the \(LM\) band (\citealt{Howarth}),  \(\lambda\) as the observing wavelength, \(B=\sqrt{u^2+v^2}\) as the baseline, \(\Theta_{\star}=2.22\pm 0.02\) mas as the stellar diameter of Fomalhaut (\citealt{Absil}) and \(J_{1}(z), J_{1.5}(z)\) as the Bessel functions. The combined visibility with contributions from the limb-darkened photosphere and the circumstellar radiation, \(V_{2}\), is given as (\citealt{Di Folco}) \begin{equation}\label{all} V_{2}(B,\lambda)=\left(1-f_{2}\right)V_{\star}(B,\lambda)+f_{2} V_{\mathrm{Gaussian}}(B,\lambda), \end{equation}
with the visibility of the circumstellar radiation (considering the symmetric Gaussian response function of \mbox{MATISSE})
\begin{equation}\label{gauss} V_{\mathrm{Gaussian}}(B,\lambda)=\exp\left({-\frac{z_{\mathrm{FOV}}^2}{4 \ln 2}}\right),\ z_{\mathrm{FOV}}=\frac{\pi\Theta_{{\mathrm{FOV}}}B}{\lambda} ,\end{equation}
with \(\Theta_{{\mathrm{FOV}}}=0.6\) as the physical FOV in the \(LM\) band. 

\subsection{One-step approach}\label{procedure2}
In contrast to the approach above (\Cref{procedure}), we avoid the inconsistency of the considered model geometry between the steps one and two by applying the following self-consistent one-step fitting approach as the main fitting approach for this study to restrict the HEZD parameters around Fomalhaut. A similar approach was introduced for the Zodiacal cloud model by \citet{Kelsall1998} and was also applied for a ring HEZD model by \citet{Absil} and \citet{Defrere}. We fit model parameters of a geometrically and optically thin dust ring or a spherical shell surrounding the star represented by a limb-darkened photosphere model with fixed parameters directly to the measured visibilities. In the following sections, the quantity \(f_{1}\)  denotes the flux ratio between the integrated flux of the ring/spherical shell and the stellar photospheric emission \(F_{\star}\). Moreover, we applied the Van-Cittert-Zernike theorem to calculate the combined visibility of the dust ring or the spherical shell and the limb-darkened photosphere (\(V_{1}\)) from the corresponding brightness distribution  \(I_{1}\), calculated with the simulation tool debris disks around main-sequence stars (DMS, \citealt{DMS}; \citealt{Stuber}):
\begin{equation}\label{Ring} V_{1}(B,\lambda)=\frac{\int_{-\infty}^{\infty} \int_{-\infty}^{\infty} I_{1}(x,y,\lambda)\exp{\left(-2\pi i\left(ux+vy\right)\right)}\mathrm{d}x\mathrm{d}y}{\int_{-\infty}^{\infty} \int_{-\infty}^{\infty} I_{1}(x,y,\lambda)\mathrm{d}x\mathrm{d}y}.  \end{equation}
Here, \(x,y\) represent the angular coordinates on the sky. In addition to the two outlined fitting methods, we also use the bootstrapping technique (\Cref{bootstrapping}) and neural networks (see \Cref{neural}) as alternative approaches to analyze the MATISSE data.

\subsection{HEZD fitting model}\label{procedure3}
For the sake of conformity, we used a similar ring model as in previous HEZD studies but with more free parameters and larger parameter ranges. Our ring model is defined by an inner radius \(R_{\mathrm{in}}\), an outer radius \(R_{\mathrm{out}}\), a ring inclination \(i\) (\(i=0^\circ\) is face-on orientation), a steeply descending number density \(n(r)=r^{-4}\) of the dust with \(r\) as the radial distance to the star and a HEZD mass \(\rm{M}\). We use three different dust species: amorphous carbon (\citealt{Rouleau}), astronomical silicate (\citealt{Weingartner}), and graphite with the standard 1/3-2/3 approximation for the calculation of the optical properties (\citealt{Draine}). The dust grains are assumed to be compact spheres with a bulk density of  \(\rho = 2.24\ \text{g}\ \text{cm}^{-3}\) for graphite and amorphous carbon, and a bulk density of \(\rho = 3.80\ \text{g}\ \text{cm}^{-3}\) for astronomical silicate. We select a narrow size distribution with a resolution of \(\Delta a = 0.3 a\) (\citealt{Kirchschlager2020}) in order to avoid artifacts due to resonances in the case of discrete grain sizes (e.g., \citealt{Draine}).  Unless otherwise noted, we fixed the position angle (PA) of the ring to \(156^\circ\) in agreement with the position angle of the outermost ring of Fomalhaut (\citealt{Kalas2005}). The optical properties of the dust grains are calculated with the software tool miex (\citealt{Wolf}). For the model of the spherical shell, corresponding parameters are used except for the ring inclination, which is not required.\\
\\
For each set of parameters, the goodness of the fit was evaluated by the reduced chi-squared \(\chi_{\mathrm{red}}^{2}\) (except for bootstrapping), defined as \begin{equation}\label{chi} \chi_{\mathrm{red}}^{2} = \frac{\chi^2}{\nu},\\ \chi^2 = \sum_{i} {\frac{(V_{\mathrm{obs,i}} - V_{\mathrm{calc,i}})^2}{V_{\mathrm{err,i}}^2}},\end{equation}
 with \(\nu\) as the degrees of freedom, \(V_{\mathrm{obs,i}}\) as the \(i\)-th measured visibility, \(V_{\mathrm{calc,i}}\) as the \(i\)-th calculated visibility (see \hyperref[Ring]{Eq.~\ref*{Ring}}, \hyperref[all]{Eq.~\ref*{all}}), and \(V_{\mathrm{err,i}}\) as the measurement error of the \(i\)-th measured visibility. We denote the parameter configuration that minimizes \(\chi_{\mathrm{red}}^{2}\) as the best-fit model and summarize the evaluated parameter space in \Cref{table1}. With 72 visibility measurements considered and five free (ring) model parameters to be varied, this results in \(\nu=67\) as the number of degrees of freedom. Furthermore, parameter configurations that lead to a dust temperature above the sublimation temperature of the dust species selected (2000 K for graphite and amorphous carbon, 1200 K for silicate, \citealt{Kobayashi}) are considered as not suitable to reproduce the measurements.\\
Our approach to obtain the error range of a fit HEZD parameter (denoted here with \(n\)) is based on an implementation of the maximum-likelihood method from \citet{Mennesson2014} that was also applied in a simplified form in other HEZD studies (e.g., \citealt{Kirchschlager2020}).
We defined the upper error \(\Delta n \) such that if \(n'\) deviates from the best-fit parameter \(n_{\mathrm{best-fit}}\) by \(\Delta n \), the probability distribution  \(p(\chi_{\mathrm{red}}^{2})=\exp\left(-\chi_{\mathrm{red}}^{2}+\chi_{\mathrm{best-fit}}^{2}\right)\) is halved, i.e.
\begin{equation} \label{interval} n'=n_{\mathrm{best-fit}}+ \Delta n \rightarrow \chi_{\mathrm{red}}^{2} = \chi_{\mathrm{best-fit}}^{2}+ \ln{2} \rightarrow p(\chi_{\mathrm{red}}^{2})=0.5.   \end{equation}
In analogy, the lower error interval  is defined as \(n''=n_{\mathrm{best-fit}}- \Delta n'\).

\begin{table*}[t!]
\caption{Evaluated parameter space  applying the one-step approach for the ring model and the spherical shell model (both for the MATISSE data).}
\label{table1}
\centering
\begin{tabular}{c c c} 
 \hline
 Parameter & Explored range & Number of linearly distributed values \\ 
 \hline
 Inner ring/shell radius \(R_{\mathrm{in}}\) [au] & 0.01 \(-\) 8 & 10000 \\ 
 Outer ring/shell radius \(R_{\mathrm{out}}\) [au] & \(0.011-60\)& 1000 \\
 Dust grain radius  \(\em a\) [\(\mu\)m] &\(0.01-10\) & 10000 \\
 Ring inclination \(i\) [\(^\circ\)] & \(0-90\) & 100 \\
 Total dust mass M  [\(\rm{M}_{\oplus}\)] & \(10^{-10}-10^{-8}\)  & 10000     \\
 
 \hline
\end{tabular}

\end{table*}

\section{Results}\label{results}
First, we investigate whether a detection of circumstellar radiation around Fomalhaut has occurred by analyzing the MATISSE data in \Cref{analysis}. 
We investigate in \Cref{planetary} whether and if so for which parameter values a stellar companion around Fomalhaut can reproduce the MATISSE data. Subsequently, in \Cref{bestfit} we specify which HEZD parameter configuration reproduces the MATISSE data best, applying the one-step approach. We also use inferred fluxes from previous Fomalhaut studies (VINCI + KIN data) aiming to further restrict the HEZD properties. Then, we also apply a double-ring model similar to a previous Fomalhaut study (\citealt{Lebreton}) to constrain the HEZD parameters for the \mbox{MATISSE}+VINCI+KIN data using the one-step approach in \Cref{previous}. Moreover, as alternative fitting approaches, we apply the bootstrapping technique in \Cref{bootstrapping} and neural networks in \Cref{neural}. Finally, we compare the different dust-to-star flux ratios resulting from the applied fitting approaches considering the MATISSE data in \Cref{steps}.
\subsection{Data analysis applying the two-step approach}\label{analysis}
To investigate the significance of the MATISSE detection of circumstellar radiation we applied the first step of the two-step approach (see \Cref{procedure}) to calculate the best-fit dust-to-star flux ratio (see \Cref{table10} for the flux ratios \(f_{2}\) and the maximum errors \(\Delta f_{2}\)) at each observing wavelength. Considering the \(3\sigma\) condition \((f_{2}/|\Delta f_{2}|\geq3\)) as a requirement for a significant detection, there is no significant detection of circumstellar radiation in the \(LM\) band. For a wavelength of \(3.06\ \mu\)m, there is almost a \(1\sigma\) detection. However, we note that for the limb-darkened stellar photosphere surrounded by a uniform circumstellar radiation the corresponding best-fit \(\chi_{\mathrm{red}}^{2}\) (see \hyperref[chi]{Eq.~\ref*{chi}} for the definition, values in the range of \(0.07-0.33\)) is only at observing wavelengths in the range of \(3.06 - 3.54\ \mu\mathrm{m}\) marginally lower than the best-fit \(\chi_{\mathrm{red}}^{2}\) of the  visibility of the limb-darkened photosphere (values in the range of \(0.07-0.66\)). This points toward potential circumstellar radiation around Fomalhaut within the FOV of \mbox{MATISSE}, though additional data are required to confirm this.  At the observing wavelengths in the range of \(3.65 - 4.91\ \mu\mathrm{m}\), a bare limb-darkened photosphere provides either the same or even a better fit to the measurement data than  circumstellar radiation, due to the larger measurement errors especially for the \(M\) band data (see again \Cref{Fig first}). However, even this non-detection (considering the  \(3\sigma\) condition) in this previously not explored wavelength range has the potential to provide tight constraints for the model parameters of the HEZD around Fomalhaut. Furthermore, although the measured closure phases for the four triangles of the observation in the \(L\) band are all close to \(0^\circ\) (see again \Cref{Fig 3}) and they also show no signal at low amplitude, a stellar companion cannot be ruled out (see \Cref{planetary}). 

\begin{figure}[t!]
\begin{minipage}{0.44\textwidth}
\includegraphics[width=\textwidth]{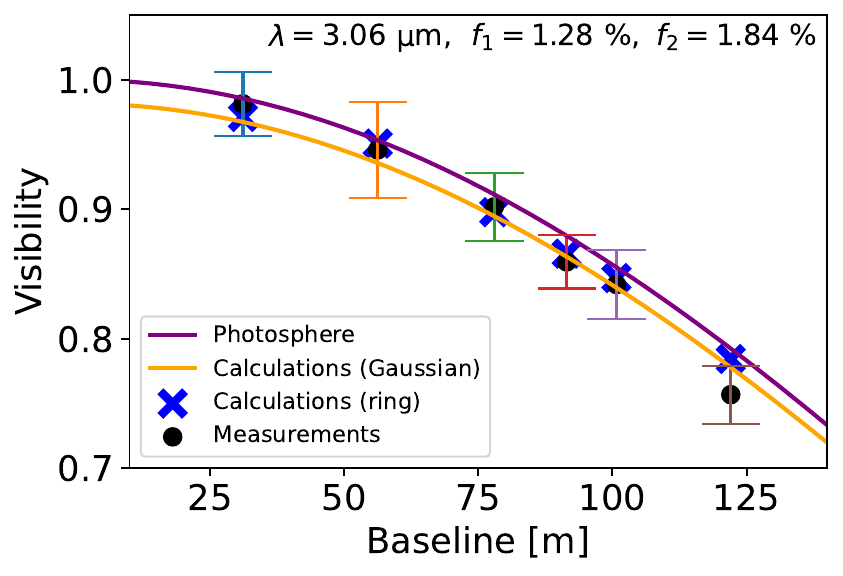}

\end{minipage}
\begin{minipage}{0.44\textwidth}
\includegraphics[width=\textwidth]{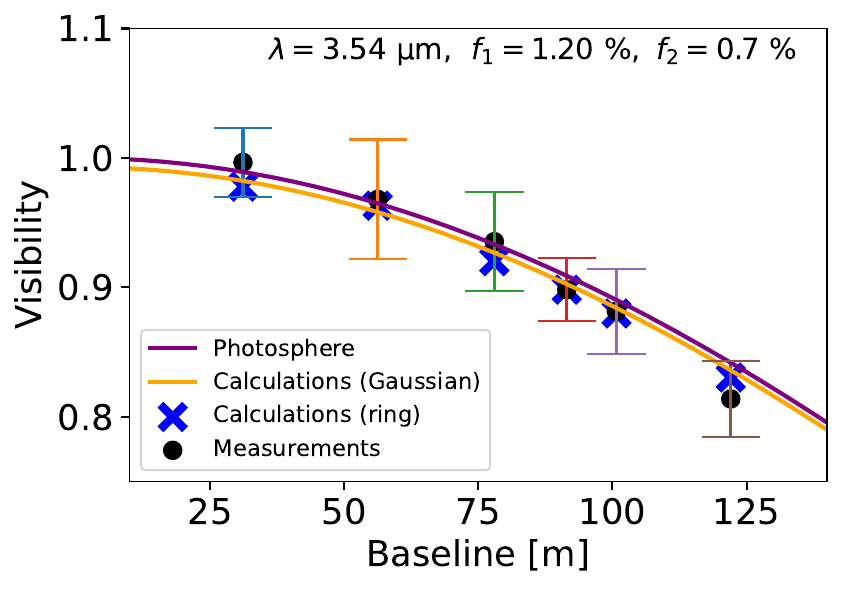}

\end{minipage}
\begin{minipage}{0.44\textwidth}
\includegraphics[width=\textwidth]{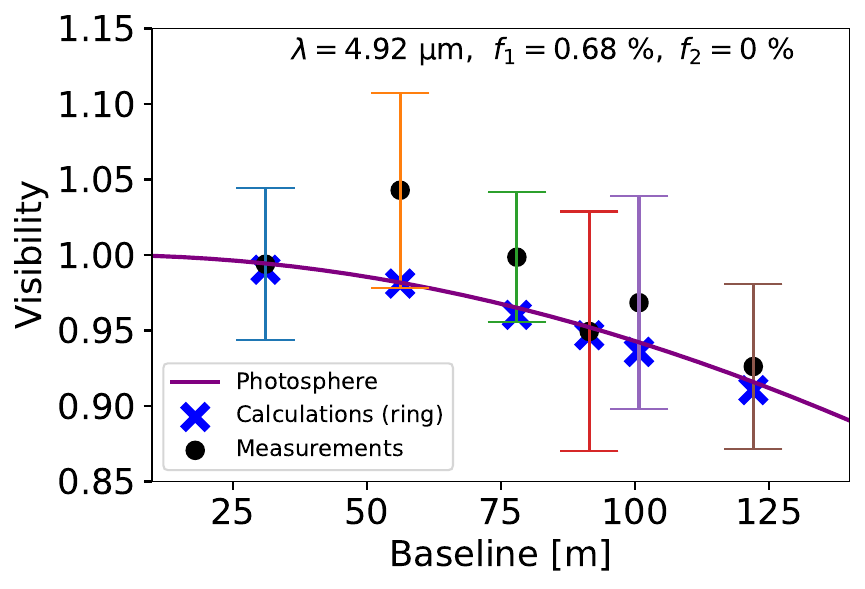}

\end{minipage}
\caption{Visibility of the limb-darkened photosphere, best-fit ring model from the one-step approach, uniform circumstellar radiation from the two-step approach (Gaussian response function of \mbox{MATISSE}), and the calibrated MATISSE data dependent on the baseline at three representative wavelengths. Top: \(3.06\ \mu\mathrm{m}\ (L\ \mathrm{band})\). Mid: \(3.54\ \mu\mathrm{m}\ (L\ \mathrm{band})\). Bottom: \(4.92\ \mu\mathrm{m}\ (M\ \mathrm{band}\), see \Cref{analysis} and \Cref{bestfit} for details).} \label{Fig 2}
\end{figure}

\subsection{Constraints on a stellar companion}\label{planetary}
To evaluate whether the presence of a binary companion could explain the obtained MATISSE data, a model was constructed under the assumption that the companion orbits within the plane of the dust ring. The projected orbit was defined with an inclination of \(65.6^\circ\) (orbital inclination of Fomalhaut's outermost ring) and a position angle of \(156^\circ\). According to a similar approach from \citet{Absil} for VINCI data, a circular orbit was assumed, reducing the model parameters to the semi-major axis, the orbital phase at a reference time, and the binary contrast between the companion and Fomalhaut.
Using a selected range of values for these parameters, the resulting visibilities and closure phases at every observing wavelength were computed. For each combination of semi-major axis and binary contrast, the orbital phase was optimized to minimize the \(\chi_{\mathrm{red}}^{2}\) between the observed and simulated visibilities, respectively closure phases. The analysis focused on binary contrasts below 5\% considering the \(LM\) band data, as brighter companions would have been presumably detected with spectroscopic observations for this observing wavelength range.\\
\\
Considering a semi-major axis in the range of \(0.05 - 10\) au, we find a best-fit  \(\chi_{\mathrm{red}}^{2}\) of 0.41 for a semi-major axis of 0.32 au and a binary contrast of 2.1 \% for the visibility fitting and a best-fit \(\chi_{\mathrm{red}}^{2}\) of 0.51 for a semi-major axis of 0.34 au and a binary contrast of 2.9 \% for the closure phases fitting. However, no error intervals can be determined for either the semi-major axis or the binary contrast, so that even for the outermost semi-major axes and thus low binary contrasts, a sufficiently good fit to the measured values can be determined. Thus, neither the semi-major axis nor the binary contrast of a possible companion can be constrained.
In conclusion, the MATISSE data alone cannot definitively confirm or rule out the presence of such a binary companion.
\subsection{Constraints on the HEZD parameters applying the one-step approach}\label{bestfit}
 \begin{table*}[t!]
\caption{Overview of the best-fit dust-to-star flux ratios derived from different modeling and fitting approaches for the MATISSE data.}
\label{table10}
\centering
\begin{tabular}{c c c c c c c c c c} 
 \hline
 Wavelength [\(\mu\)m] & \(f_{1}\) [\%] & \(f_{2}\) [\%] & \(\Delta f_{2}\) [\%] & \(f_{\mathrm{Ring}}\) [\%] & \(\Delta f_{\mathrm{Ring}}\) [\%] & \(f_{\mathrm{Gaussian}}\) [\%] & \(\Delta f_{\mathrm{Gaussian}}\) [\%] & \(f_{\mathrm{NN}}\) [\%] & \(F_{\star}\) [Jy] \\ 
 \hline
 3.06 & 1.28 & 1.84 & 2.16 & 2.42 & 0.91 & 2.02 & 0.60 & 0.90 & 173.93 \\ 
 3.18 & 1.30 & 1.56 & 2.28 & 2.16 & 1.05 & 1.84 & 0.77 & 0.90 & 162.93 \\ 
 3.30 & 1.28 & 1.42 & 2.31 & 1.81 & 0.80 & 1.79 & 0.97 & 0.87 & 152.87 \\ 
 3.42 & 1.24 & 1.12 & 2.40 & 1.41 & 0.88 & 1.34 & 0.76 & 0.85 & 143.78 \\ 
 3.54 & 1.20 & 0.70 & 2.62 & 1.02 & 0.75 & 0.81 & 0.62 & 0.83 & 135.40 \\ 
 3.65 & 1.14 & 0.20 & 2.77 & 0.36 & 0.80 & 0.22 & 0.72 & 0.80 & 127.73 \\ 
 3.78 & 1.07 & 0    & 2.61 & 0    & 0.68 & 0    & 0.48 & 0.78 & 120.67 \\ 
 3.89 & 1.02 & 0    & 2.54 & 0    & 0.60 & 0    & 0.42 & 0.75 & 114.20 \\ 
 4.55 & 0.78 & 0    & 3.23 & -    & -    & -    & -    & 0.64 & 86.31  \\ 
 4.68 & 0.74 & 0    & 2.72 & -    & -    & -    & -    & 0.62 & 82.35  \\ 
 4.80 & 0.71 & 0    & 2.80 & -    & -    & -    & -    & 0.61 & 78.65  \\ 
 4.91 & 0.68 & 0    & 2.74 & -    & -    & -    & -    & 0.59 & 75.20  \\ 
 \hline
\end{tabular}
\tablefoot{
 The best-fit dust-to-star flux ratios are denoted according to the applied fitting approach: One-Step (\(f_{1}\)), Two-Step (\(f_{2}\)), Bootstrapping (\(f_{\text{Ring}}, f_{\text{Gaussian}}\)), and neural networks (\(f_{\text{NN}}\)). ``-'' indicates that the bootstrapping technique was not applied at these wavelengths. The last column indicates the stellar flux \(F_{\star}\).
}
\end{table*}

\begin{figure}[t!]
\includegraphics[scale=0.21]{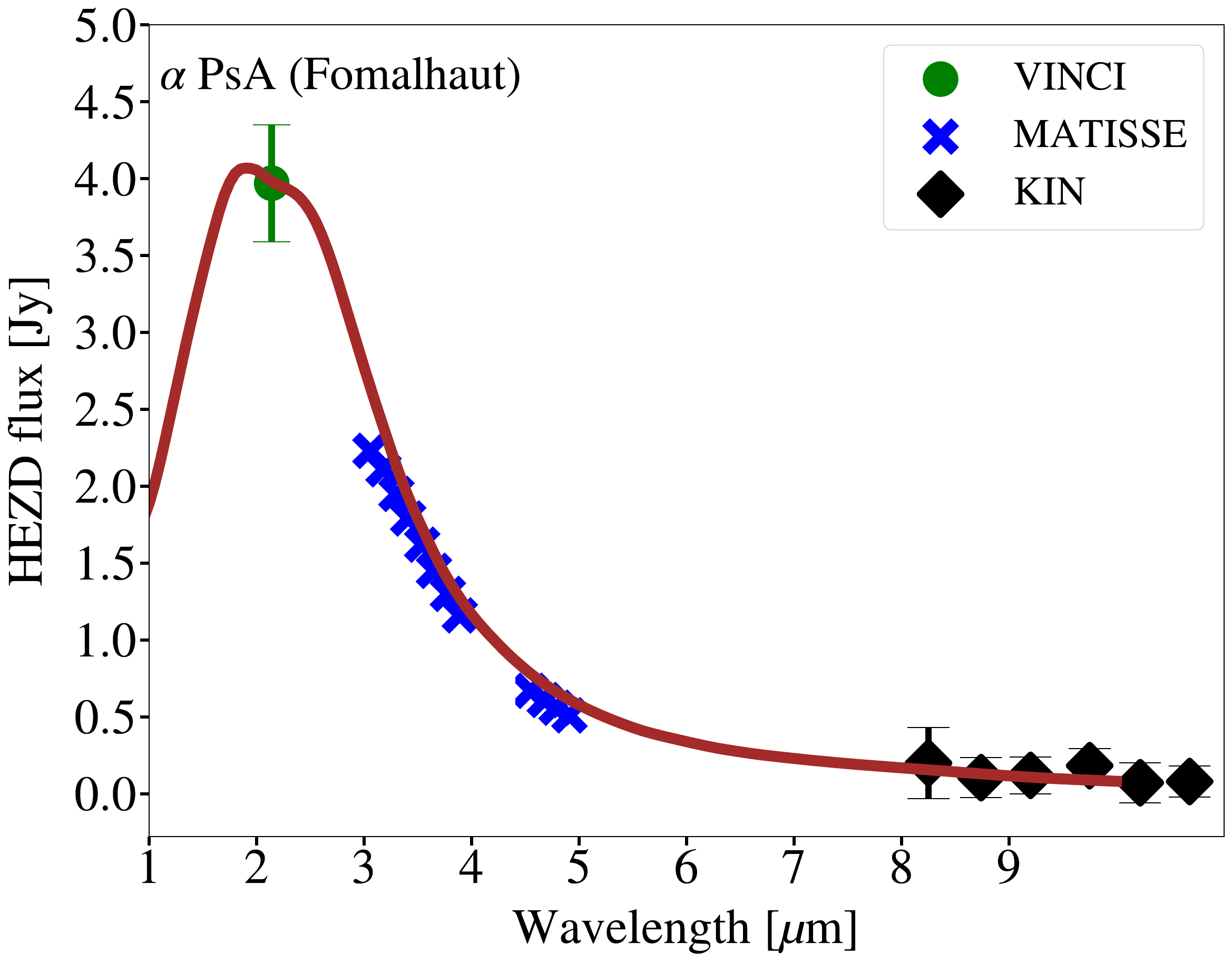}
\caption{Spectral energy distribution  of the HEZD inferred from \(K, L, M,\) and \(N\) band observations with a  best-fit parameter configuration for the ring model (see \Cref{table5} and \Cref{bestfit} for details).}\label{Fig 5}
\end{figure}

 Using the obtained \(LM\) band MATISSE data, we try to follow up the previous Fomalhaut HEZD studies and aim to characterize the circumstellar radiation around Fomalhaut more precisely applying the one-step fitting approach for the whole measurement data (and thus every observing wavelength) simultaneously. Considering \Cref{Fig 2}, at two of the three example observing wavelengths, \(\lambda=3.06\ \mu\)m and \(\lambda=3.54\ \mu\)m, a marginal visibility deficit to the expected visibility of Fomalhaut's photosphere is recognizable (see \Cref{analysis} for this discussion). The visibilities for the best-fit ring model and the best-fit uniform circumstellar radiation model show no significant difference considering their goodness of fitting the \(L\) band data and there is a marginal deviation to the goodness of the fit with a bare limb-darkened photosphere (\(\chi_{\mathrm{red}}^{2}=0.26\)). Regarding the best-fit of the ring model  (\(\chi_{\mathrm{best-fit}}^{2}=0.23\)) and the spherical shell model (\(\chi_{\mathrm{best-fit}}^{2}=0.25\)), only marginal differences between the values of the best-fit parameters are noticeable (see \Cref{table5}). Without loss of generality, we thus refer to the ring model in the following. The best-fit parameters are also consistent with those from earlier Fomalhaut HEZD studies (see \Cref{discussion}).\\
 Although the ring inclination cannot be constrained by the obtained data, large inclinations (close to edge-on) are nevertheless preferred due to their slightly lower \(\chi_{\mathrm{red}}^{2}\) in contrast to lower inclinations (see also \Cref{neural}), which results from a small contribution of scattered radiation for these larger inclinations.\\
\\
The calculated best-fit parameters are further supported by both the Akaike information criterion (AIC, \citealt{Akaike}) and the Bayesian information criterion (BIC, \citealt{Schwarz}). These two criteria are used to compare two models and determine which one fits the measurement data better (lower AIC, respectively lower BIC). For a ring model with the calculated best-fit parameter values, an AIC of -390 and a BIC of -386, and for a bare limb-darkened photosphere, an AIC of -387 and a BIC of -383 is determined. The significance criterion in the context of these criteria only applies to a difference of about five between the AIC (or BIC) of two models, so no significant detection of HEZD can be determined with these information criteria either for our considered data set.\\
\\
Considering the error intervals for the different HEZD parameters, we further recognize that the MATISSE data are not sufficient to constrain the HEZD parameter values (see \Cref{table2}, error intervals do not exist considering only the MATISSE data).
With the aim to pose tighter constraints on the HEZD properties, we used complementary VINCI and KIN data from \citet{Absil} and \citet{Mennesson2014}. These studies inferred the radiation emanating from the circumstellar environment in comparison to the sole contribution of the central star. Using the corresponding HEZD fluxes (3.97 \(\pm\) 0.38 Jy for a dust-to-star flux ratio of \(1.26\%\) at an effective wavelength of 2.12\ \(\mu\)m (\citealt{Absil}) and six HEZD fluxes at six wavelengths in the range of \(8.25 - 10.68\ \mu\mathrm{m}\) (\citealt{Mennesson}), we applied spectral energy distribution (SED) fitting for the inferred HEZD fluxes (see \Cref{Fig 5} for the resulting SED, we calculated the HEZD fluxes with DMS) in combination with the one-step approach for the MATISSE data. Considering the inferred fluxes from the KIN measurements, we calculate the HEZD fluxes for each observing wavelength using \(F_{\mathrm{HEZD}}=2.5 F_{\star}E_{\lambda}\) (\citealt{Kirchschlager2017}) where \(E_{\lambda}\) is the measured excess leak given in the third segment of Table 3 from \citet{Mennesson}. In this context, we have taken into account the inner working angle (6 mas) and outer working angle (200 mas) of KIN.\\
The inner radius is limited within the error interval in the range of  \(0.10 - 0.17\) au and the dust grain radius is constrained by \(0.64\ \mu\mathrm{m}\) (see \Cref{table2}). Thus, the origin of the circumstellar radiation around Fomalhaut is within the FOV of MATISSE (\(\sim\) 1 au). We note that the theoretical blow-out size of dust grains around Fomalhaut amounts to \(\sim 3.0\ \mu\)m (\citealt{Kirchschlager2013}) and smaller grains should be blown out of the system. The corresponding dust temperatures are between 920 K and 1995 K, whereby the upper temperature limit is close to the sublimation temperature of amorphous carbon (2000 K). However, the potential corresponding trapping or delivery mechanisms for those hot dust grains are beyond the scope of this study.\\
By combining the MATISSE with the KIN and VINCI measurements, we neglect the possible temporal variability of the HEZD radiation. We note that \(\kappa\) Tuc is so far the only investigated HEZD system that has shown clear temporal variability in the \(H\) band (\citealt{Ertel2016}). We conclude that while the MATISSE data in combination with the VINCI+KIN measurements do not provide any additional HEZD parameter constraints compared to an exclusive consideration of the VINCI+KIN measurements alone, they are nevertheless consistent with the model parameter constraints derived in these previous studies.

\begin{table*}[t!]
\caption{Summary of the best-fit parameters for fitting the MATISSE data with different models  with amorphous carbon as the best-fit dust species.}
\label{table5}
\centering
\resizebox{\textwidth}{!}{%
\begin{tabular}{lccccc}
\hline
Parameter & Ring (one-step) & Ring (NN) & Spherical Shell  & Double Ring (\citealt{Lebreton})  & Double Ring  \\
\hline
\(\chi_{\mathrm{red}}^2\) & 0.23 & 0.22 & 0.25 & 1.60 & 0.51 \\
Inner radius \(R_{\mathrm{in}}\) [au] & 0.11 & 0.2 & 0.12 & 0.09 / 1.59 & 0.12 / 2.39 \\
Outer radius \(R_{\mathrm{out}}\) [au] & 0.12 & 0.202 & 0.13 & 0.10 / 2.10 & 0.14 / 3.43 \\
Grain size \(a\) [\(\mu\)m] & 0.53 & 0.48 & 0.61 & 0.01–1000 / 3.51–1000 & 0.41 / 47 \\
Inclination \(i\) [\(^\circ\)] & 90 & 90 & – & 65.6 / 65.6 & 65.6 / 65.6 \\
Total dust mass \(M\) [\(M_\oplus\)] & \(3.25 \times 10^{-10}\) & \(5.35 \times 10^{-10}\) & \(3.61 \times 10^{-10}\) & \(2.50 \times 10^{-10}\) / \(2.86 \times 10^{-6}\) & \(4.04 \times 10^{-10}\) / \(5.79 \times 10^{-7}\) \\
\hline
\end{tabular}
}
\tablefoot{
 For the double-ring models, the inner and outer ring components are shown separately and distinguished with "/". The inner ring is composed of amorphous carbon; the outer ring consists of a 50:50 mixture of astronomical silicates and amorphous carbon. For the double-ring model from \citet{Lebreton} consider Table 2 (free outer slope, VINCI+KIN data) from the corresponding publication. For our double ring-model we considered the MATISSE+VINCI+KIN data.
}
\end{table*}

\begin{table*}[t!]
\caption{Parameter space for the error interval applying the one-step approach (ring model, amorphous carbon).}
\label{table2}
\centering
\begin{tabular}{c c} 
 \hline
 Parameter & Error interval (MATISSE+VINCI+KIN) \\ 
 \hline
 Inner ring radius \(R_{\mathrm{in}}\) [au] & 0.10 \(-\) 0.17 \\ 
 Outer ring radius \(R_{\mathrm{out}}\) [au] & 0.11 \(-\) 0.32 \\
 Dust grain radius  \(\em a\) [\(\mu\)m] &\(\leq0.64\) \\
 Ring inclination \(i\) [\(^\circ\)] & \(0-90\) \\
  Total dust mass \(\rm{M}\) [\(\rm{M}_{\oplus}\)] &  \(1.21\times 10^{-10} - 3.43\times 10^{-10}\)  \\
 \hline
\end{tabular}
\tablefoot{
 The error interval for the HEZD parameters referring to the VINCI+KIN data is the same as for the MATISSE+VINCI+KIN measurements.
}
\end{table*}

\subsection{Constraints from the MATISSE observation in context of a double-ring model}\label{previous}

The MIR measurements with KIN  in combination with the NIR measurements with VINCI at \(2.12\ \mu\)m are consistent with the existence of two distinct dust populations within a few au around Fomalhaut (\citealt{Lebreton}; \citealt{Mennesson}). Those studies concluded on a population of very small and sub-blowout, hot (\(\sim\) 2000 K) dust grains confined in a narrow region (\(0.1 - 0.3\) au) at the sublimation rim of carbonaceous material and a population of larger, warm (\(\sim\) 400 K) bound grains at \(\sim\) 2 au (outer ring), applying a steep grain size distribution and a flatter dust density distribution. The outer ring was assumed to be made of a  half mixture of astronomical silicate and carbonaceous material. We note that the ring inclination in this model was not a free parameter but was chosen to match the inclination of the outer disk around Fomalhaut (\(65.6^\circ\), see \Cref{table5} for an overview about the best-fit parameter values of \citeauthor{Lebreton} \citeyear{Lebreton}). Moreover, the maximum grain radius was set to \(1000\ \mu\)m because it cannot be constrained by the adopted modeling approach.\\
\\
Our best-fit parameter values for the inner ring of a double-ring model do not differ significantly from our results considering the single-ring model (see \Cref{table5}, \(\chi_{\mathrm{best-fit}}^{2}=0.51\) for our best-fit parameter values of the double-ring model) for the MATISSE+VINCI+KIN data. Despite the mass of the outer ring being two orders of magnitude higher than the mass of the inner ring, the best-fit \(\chi_{\mathrm{red}}^{2}\) for the double-ring is nevertheless similar to that of the single-ring model. The differences between the best-fit parameter values of \citet{Lebreton} and our results are based on the circumstances that our double-ring model has different assumptions for the dust particle size distribution (\(\Delta a = 0.3 a\)) and the dust density distribution (\(n(r)=r^{-4}\)) respectively for the inner and the outer ring as outlined in \Cref{model}. Noticeable are the significantly larger dust grains of which the outer ring consists (\(\em\)\( a=47\ \mu\mathrm{m}\)) compared to the dust grains of which the inner ring consists (\(\em\)\( a=0.41\ \mu\mathrm{m}\)), whereby we assumed the same inclination for both rings (\(65.6^\circ\)) for a proper comparison with the results from \citet{Lebreton}. Moreover, we also used a different concept of sublimation: While \citet{Lebreton} introduces a time span during which a dust particle above the sublimation temperature can still be present and be evaluated as a valid dust particle for the HEZD model, we immediately exclude all dust particles from our model that exceed the sublimation temperature, which results in different best-fit parameter values. Since the visibilities resulting from the best-fit parameter values of the inner and outer ring do not fit the MATISSE+VINCI+KIN data significantly better than the visibilities resulting from best-fit parameter values from a single-ring model, we conclude that a double-ring model is not more likely implied by an analysis of the MATISSE data than a single-ring model.

\subsection{Data analysis applying bootstrapping} \label{bootstrapping}

 As an alternative approach to investigate the possible presence of circumstellar radiation around Fomalhaut instead of using the \(\chi_{\mathrm{red}}^{2}\) method, a bootstrapping algorithm is applied as it was done in previous HEZD studies (e.g., \citealt{Ertel}; \citealt{Absil2021}). Details of the bootstrapping technique are provided in \Cref{bootstrappingmethod}.\\
 Here, \(f_{\mathrm{Gaussian}}\) denotes the best-fit flux ratio between the uniform circumstellar radiation (Gaussian response function of MATISSE) around Fomalhaut and the emission of the limb-darkened photosphere (see \hyperref[gauss]{Eq.~\ref*{gauss}} for the visibility function). Besides, \(f_{\mathrm{Ring}}\) denotes the best-fit flux ratio between the emitted and scattered flux of the dust ring and the emission of the limb-darkened photosphere. To determine \(f_{\mathrm{Ring}}\) with bootstrapping, the inner and outer ring radius were considered as free parameters using

 \begin{equation}\label{ringanalytical}
V_{\mathrm{Ring}}(B,\lambda) = \left| \frac{2}{1 - \left(\frac{R_{\mathrm{in}}}{R_{\mathrm{out}}}\right)^{2}}
 \left( \frac{ J_{1}(k) - \frac{R_{\mathrm{in}}}{R_{\mathrm{out}}} J_{1} \left(\frac{k}{\frac{R_{\mathrm{out}}}{R_{\mathrm{in}}}}\right) }{k} \right) \right|,
\ k=\frac{2\pi R_{\mathrm{out}}B}{\lambda}\end{equation} as the visibility function of the ring (\citealt{Born}). An overview about the best-fit flux ratios and their maximum errors is again given in \Cref{table10}. While the inner radius has a best-fit value of \(0.18 - 0.41\) au depending on the observing wavelength, the outer radius of the ring has a best-fit value of \(0.61 - 2.13\) au.
Considering the ring model, up to \(2\sigma\)  flux ratios were identified at wavelengths of  \(3.06\ \mu\mathrm{m}\), \(3.18\ \mu\mathrm{m}\), and \(3.30\ \mu\mathrm{m}\). For the Gaussian model even a \(3\sigma\)  significant flux ratio at \(3.06\ \mu\mathrm{m}\) was identified. While most flux ratios do not show a significant detection (\(3\sigma\) condition), some values approach at least the \(1\sigma\) threshold  and reveal a consistent trend of decreasing flux ratio and significance with increasing wavelength, a pattern that aligns with findings obtained through the \(\chi_{\mathrm{red}}^{2}\) method.
\subsection{Data analysis using neural networks}\label{neural}
\begin{figure}[t!]
   \centering
   \includegraphics[width=0.94\hsize]{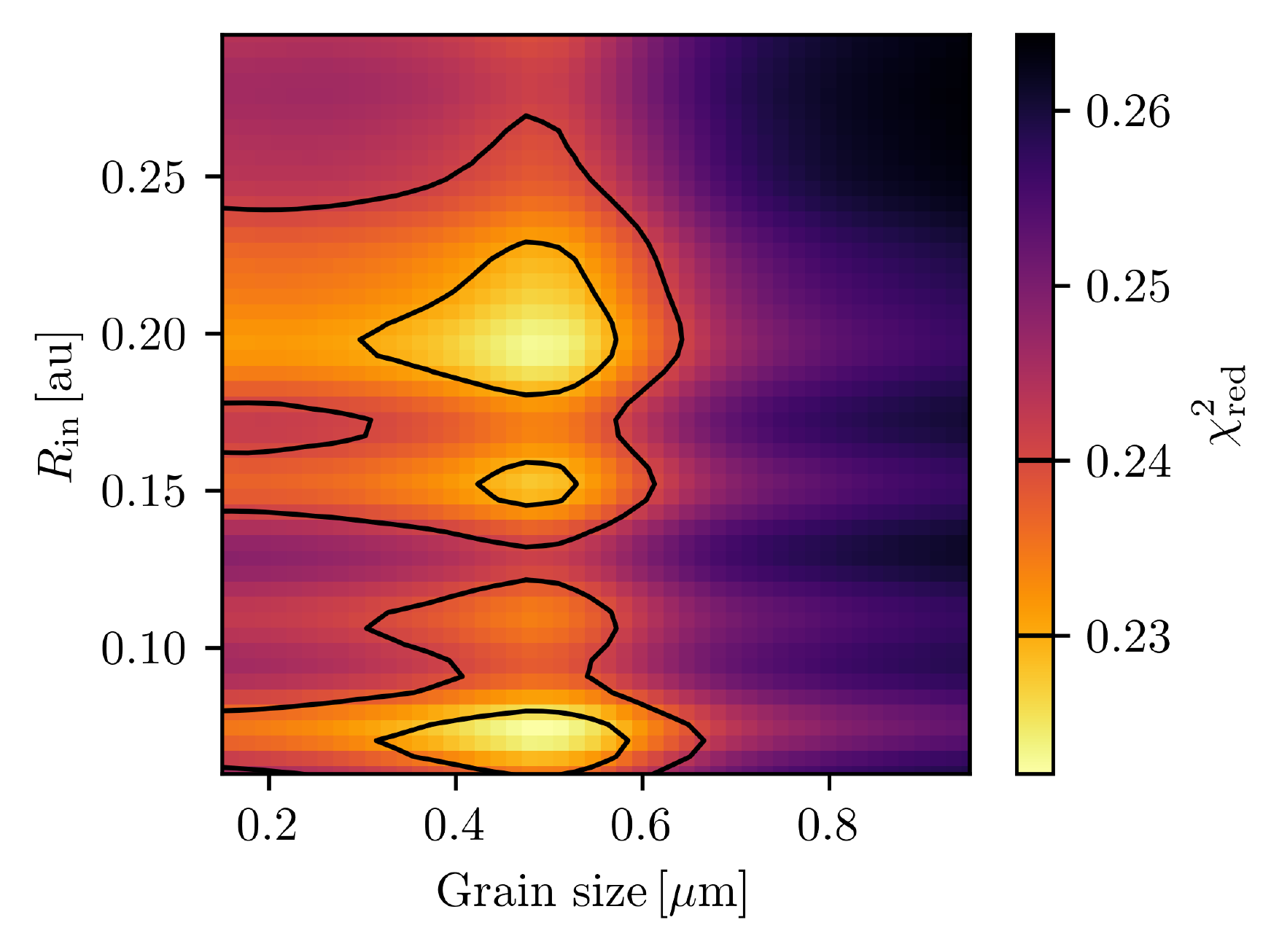}
      \caption{Constructed $\chi^2_{\text{red}}$-map for a relative width of $R_{\text{out}}/R_{\text{in}}=1.01$  varying the inner ring radii and dust grain radii using neural networks (see \Cref{neural} for details).}
         \label{fig:chi_map}
   \end{figure}

\begin{figure}[t!]
   \centering
   \includegraphics[width=1.04\hsize]{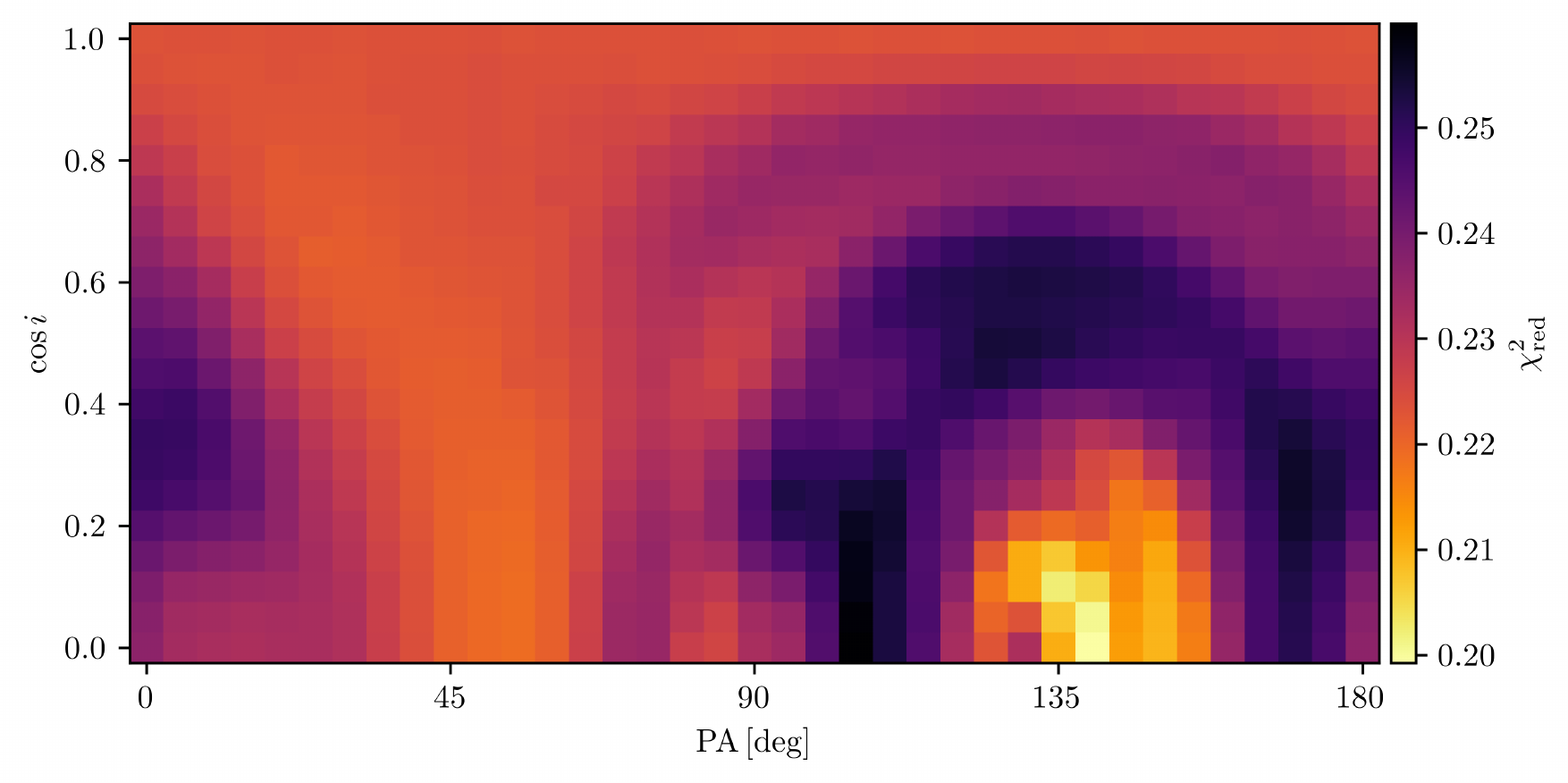}
      \caption{Constructed $\chi^2_{\text{red}}$-map for the  best-fit parameters and varying inclinations ($i$) and position angles (PA) of the ring using neural networks (see \Cref{neural} for details).
              }
         \label{fig:inc}
   \end{figure}

As another complementary approach, we use neural networks to find the best-fit HEZD parameter configuration using the \mbox{MATISSE} data, and thus $\chi^2_{\text{red}}$ was calculated for each combination of the free parameters (see \Cref{Network} for details). This procedure was performed on a grid that is specified in \Cref{tab:modell_FH}, which summarizes the investigated parameter space. Additionally, for each parameter combination we determined the disk mass (range of \(10^{-10}-10^{-8}\) \(\rm{M}_{\oplus}\)) that results in the lowest possible value of $\chi^2_{\text{red}}$, by applying a Basin-Hopping algorithm (\citealt{Wales}) to find the global minimum. \Cref{fig:chi_map} shows the resulting $\chi^2_{\text{red}}$-map for a relative width of $R_{\text{out}}/R_{\text{in}}=1.01$. The plot exhibits two local minima, both roughly at $a\approx0.5\,\mu$m. Considering only the region for $R_{\text{out}}$ larger than the sublimation radius, the best-fit with $\chi^2_{\text{red}}=0.22$ for a face-on configuration is found for the parameters $R_{\text{in}}=0.2\,$au, $R_{\text{out}}=0.202\,\text{au}$, \(\rm{M}=5.35\cdot10^{-10}\)\ \text{M}\(_{\oplus}\), and a central grain size of $a=0.48\,\mu$m. An overview about the  best-fit dust-to-star flux ratios \(f_{\mathrm{NN}}\) is again given in \Cref{table10}. Moreover, we find that none of the displayed parameter combinations led to a significant difference compared to the determined best-fit, which is in agreement with our previous findings (see \Cref{bestfit}).\\
Since the inclination $i$ and the position angle also affect the measured visibility, we analyzed their potential effect on the resulting best-fit. \Cref{fig:inc} shows a $\chi^2_{\text{red}}$-map assuming the previously found best-fit parameters, which is obtained by varying the projected inclination $\cos i$ and the position angle. The lowest $\chi^2_{\text{red}}$ is found for an inclination of $i=90^{\circ}$ and a position angle of $\text{PA}=140^{\circ}$ with a value of $\chi^2_{\text{red}}=0.20$. However, we find no significant difference between different inclinations with regard to their corresponding $\chi^2_{\text{red}}$ values. To summarize, our best-fit parameters by applying neural networks are consistent with the best-fit parameters determined in \Cref{bestfit}.
   \subsection{Flux ratios for the fitting approaches}\label{steps}
\begin{figure}[t!]
\includegraphics[scale=0.18]{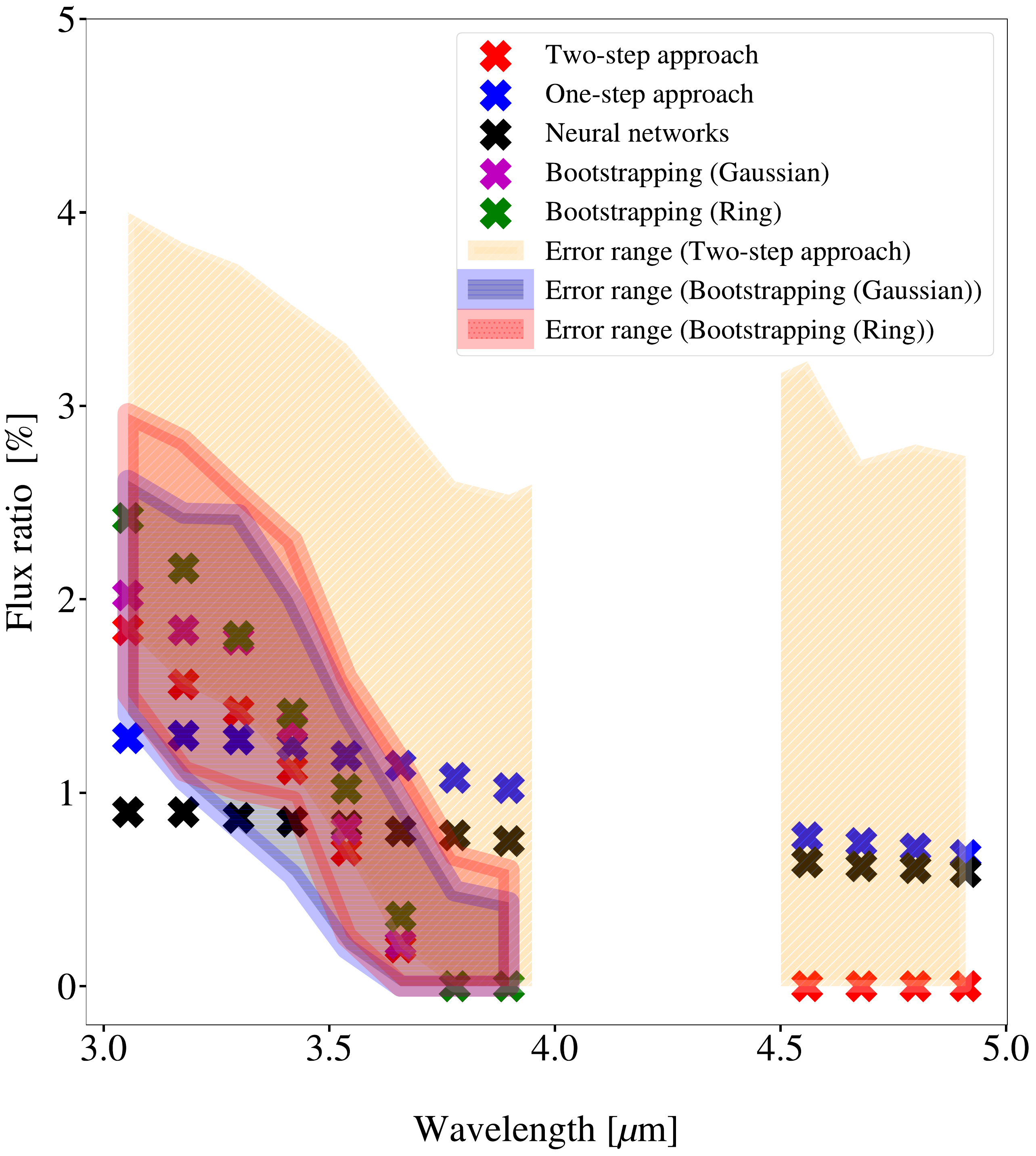}
\caption{Dust-to-star flux ratio with dependence on the wavelength with a best-fit parameter configuration applying the different fitting approaches from the previous subsections (see \Cref{steps} for details).}\label{Fig 6}
\end{figure}
Because different dust-to-star flux ratios imply different constraints for the HEZD properties, we compare the dust-to-star flux ratios for the best-fit HEZD parameter values resulting from the self-consistent one-step approach with those resulting from the other applied fitting approaches on the basis of the MATISSE observation (see again \Cref{table10}).
We observe differences at each observing wavelength for the resulting dust-to-star flux ratios and the error intervals (see \Cref{Fig 6}). While the flux ratios resulting from the two-step approach range from 0 \% to  3.9 \% (beige area), the flux ratios resulting from the one-step approach are not even restricted (no limits for the error range) so they completely overlap with those resulting from the two-step approach. This circumstance also applies for the fitting approach using the neural networks. The flux ratios resulting from the bootstrapping approaches have a tighter range;  0 \% to  2.7 \% (blue area) for the Gaussian model and 0 \% to  2.9 \% (red area) for the ring model. While the total values of the dust-to-star flux ratios differ for the various fitting methods, their trend of decreasing for increasing wavelength is the same for all of them. The flux ratios of zero arise in those fitting methods where the flux ratio itself is directly used as a fitting parameter for the measurement data. In contrast, the one-step approach and the neural network method fit a set of physical parameters describing an optically thin dust ring, from which the flux ratio is subsequently derived based on the best-fit parameter values. However, due to the low significance of our measurement data, a precise differentiation between the results is not possible in this study.

\section{Discussion}\label{discussion}
We place our fitting results in the context of previous Fomalhaut studies in \Cref{relation}. Subsequently, we discuss whether certain fitting techniques and geometrical models are preferable in such an analysis in \Cref{comparison}.

\subsection{Relation to previous Fomalhaut studies}\label{relation}
So far, there are several studies about significant circumstellar radiation around Fomalhaut in the NIR and low MIR range. The first one  (\citealt{Absil}) was based on \(K\) band (2.12\ \(\mu\)m) data obtained with VINCI (\citealt{DiFolco2004}; \citealt{Lebouquin}), which implied the presence of a NIR excess (\(0.88  \pm 0.12\ \% \) for uniform circumstellar radiation, \(1.21 \ \% \) for an optically thin ring model) caused by the thermal emission from hot submicrometer-sized grains located within 6 au from Fomalhaut. The total HEZD mass would be a few \(10^{-10}\ \rm{M}_{\oplus}\). The second study was based on KIN null depth measurements (\citealt{Mennesson}) in the \(N\) band and indicates an upper limit for an excess over the level expected from the stellar photosphere. The measured null excess has a mean value of \(0.35 \pm 0.10\ \% \) between 8 \(\mu\)m  and 11 \(\mu\)m. The source of this marginal excess must be contained within 2 au of Fomalhaut according to the small FOV of KIN. A possible explanation for both excesses is that the HEZD around Fomalhaut has a more complex geometry, for instance a double-peaked structure (\citealt{Lebreton}, see \Cref{previous} for a more detailed description). Based on the VINCI and KIN data, \citet{Kirchschlager2017} derived the following constraints for the HEZD around Fomalhaut for an optically thin ring model: an inner ring radius in the range of \(0.11 - 0.2\) au, dust grains with radius \( \leq 0.29\,\mu\mathrm{m} \), and a total dust mass between \(1.2\times 10^{-10}\ \rm{M}_{\oplus}\) and \(2.6\times 10^{-10}\ \rm{M}_{\oplus}\). The most recent of these studies was by \citet{Stuber}, where a HEZD model consisting of a optically thin ring with a bimodal grain size distribution was applied. In this study, an inner ring radius in the range of \(0.10 - 0.22\) au, two grain size distributions each consisting of particles with radii in the range of \(0.08 - 0.30\ \mu\)m and \(0.01 - 0.09\ \mu\)m, and a total dust mass between \(2.2\times 10^{-10}\ \rm{M}_{\oplus}\) and \(6.1\times 10^{-8}\ \rm{M}_{\oplus}\) were derived for the best-fit parameter values.   \\
We presented the first observation of Fomalhaut in the \(LM\) band to follow up on these studies with several fitting approaches and further restrict the HEZD parameter values applying the one-step fitting approach. We found a low significant detection of circumstellar radiation for the \(L\) band data and no significant detection for the \(M\) band data, even with multiple applied fitting approaches. However, the results considering the bootstrapping technique show at least for the \(L\) band data a  slightly higher significant result with respect to the detection of circumstellar radiation around Fomalhaut than the other applied fitting approaches. That being said, the \(3\sigma\) condition for a significant detection could be fulfilled at the observing wavelength of  \(3.06\ \mu\mathrm{m}\) for the Gaussian model (see \Cref{bootstrapping}). One explanation for the low significance of the measurement are large uncertainties in the data, with little data available at the same time (see again \Cref{Fig first}).\\
Although the range of the possible HEZD parameter values (see \Cref{table2}) cannot be further narrowed down compared to the  studies mentioned earlier, two of the parameter results are still worth mentioning:
First, whether graphite, silicate or amorphous carbon is used as the dust species is almost irrelevant for the fit of the MATISSE data with the applied approaches. Only when the KIN data in the MIR range is included, pure  silicate can be excluded as a potential dust species. Second, the best-fit ring inclination \(90^\circ\)  differs from those of the outermost cold dust ring around Fomalhaut (\(65.6^\circ \); \citealt{MacGregor}). 
However, the inclination has no major influence on \(\chi_{\mathrm{red}}^{2}\) and thus on the calculated visibilities since the ring inclination is not constrained (see \Cref{table2}). The thermal reemission for submicron-sized dust grains, which is in our case almost independent from the inclination, has a significantly larger contribution on the total HEZD flux compared to the contribution of the scattered radiation supported observationally by the lack of polarized light around several HEZD targets (\citealt{Marshall2016}).\\
\\
In contrast to \citet{Absil}, who was able to tightly constrain the semi-major axis and the binary contrast to Fomalhaut of a stellar companion and, in combination with earlier observations, also likely excluded its presence as an explanation for the measured \(K\) band excess, this restriction is not possible with the new MATISSE data. Moreover, \citet{Absil2011} found with data from the Precision Integrated Optics Near-infrared Imaging ExpeRiment (PIONIER) that a potential companion of Fomalhaut is absent within the 100 mas search region. This indicates that the VINCI detected NIR excess around Fomalhaut is related to an extended source of emission (e.g., HEZD) rather than a point-like source such as a planetary companion. 
We sum up that while the MATISSE data in combination with the VINCI+KIN measurements do not provide any additional HEZD parameter constraints compared to an exclusive consideration of the VINCI+KIN measurements alone, they are nevertheless consistent with the model parameter constraints derived in these previous studies.
 More complementary MATISSE observations are thus necessary to better constrain the nature of the \(LM\) band emission around Fomalhaut and to improve the statistical significance.
 
\subsection{Comparison of models and fitting techniques}\label{comparison}

\begin{figure}[t!]
\includegraphics[scale=0.40]{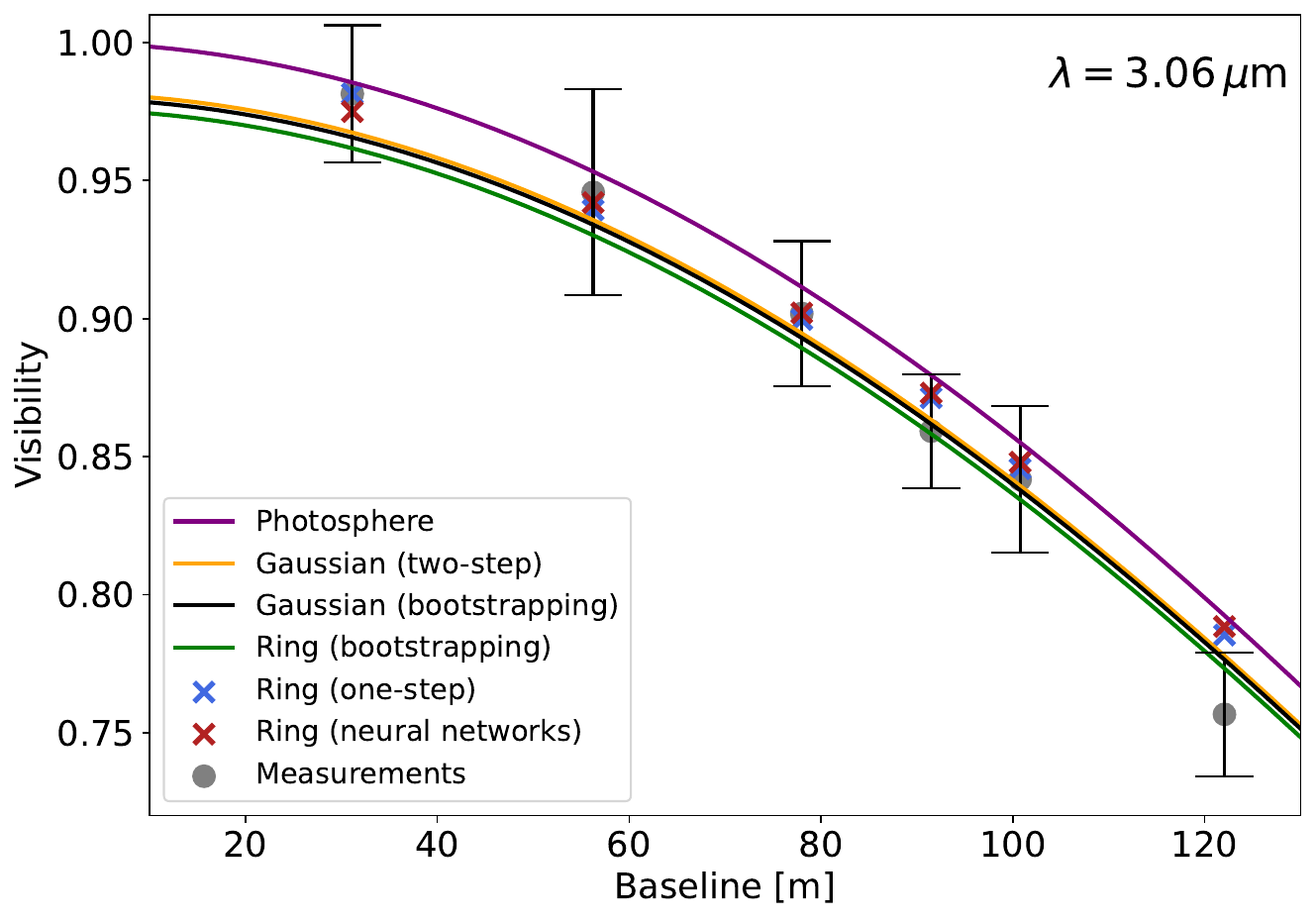}
\caption{Visibility of the limb-darkened photosphere. Visualized are all applied fitting approaches with best-fit parameter values (see \Cref{results}) and calibrated MATISSE data with dependence on the baseline at a wavelength of \(3.06\ \mu\mathrm{m}\) (see \Cref{discussion} for details).}\label{Fig 12}
\end{figure}

The selection of the geometric model and fitting method has a significant impact on both the interpretability and the computational efficiency of a HEZD analysis. Therefore, we first compare the different model geometries considered in this study, followed by a discussion of the applied fitting approaches.\\
\\
The Gaussian model  (\hyperref[gauss]{Eq.~\ref*{gauss}}) defined by an analytical function offers a simple, centrally peaked brightness distribution and is particularly well suited for statistical analyses due to its  computational efficiency. However, it provides limited interpretability (only the dust-to-star flux ratio as a free parameter), especially when modeling more  realistic dust distributions.\\
The optically thin ring model (see \Cref{procedure3})  allows for a  better representation of realistic geometries not least because of its larger number of fitting parameters. However, at the same time  degenerate solutions become more likely. Thus, this becomes particularly problematic in the case of low data quality and/or a limited \(uv\) coverage.\\ 
In contrast to the case of the optically thin ring model, the visibility function of a ring with free inner and outer radii (see \hyperref[ringanalytical]{Eq.~\ref*{ringanalytical}}) is, on the one hand, computationally efficient—similar to the Gaussian model—but, on the other hand, does not incorporate physical key parameters such as inclination, position angle, dust density distribution, grain size, or dust mass.\\
The \(\chi_{\mathrm{red}}^{2}\) method (see \hyperref[chi]{Eq.~\ref*{chi}}) provides an efficient approach for parameter estimations, yielding best-fit parameter values. However, it does not account for statistical uncertainties beyond measurement errors, which can lead to misleading results in the case of degenerate parameter spaces or overfitting, i.e., an insufficient amount of significant data relative to the number of free parameters.\\
In the two-step approach (see \Cref{procedure}) one first fits the dust-to-star flux ratios using a simple analytical model (e.g., Gaussian), followed by parameter fitting based on a fixed ring geometry. Due to its low computational effort, this method is well suited for preliminary analyses. However, it does not propagate uncertainties from the underlying geometric model fit into the final estimated parameter values of the ring model, thereby limiting its physical reliability.\\
In contrast, the one-step approach (see \Cref{procedure2}) employs the optically thin ring model by generating brightness maps, applying Fourier transforms, and directly comparing the resulting visibilities with the observational data applying the  \(\chi_{\mathrm{red}}^{2}\) method. While this method offers the highest degree of self-consistency and physical robustness—particularly when applied to extensive model grids—it is significantly more computationally demanding than the two-step approach. Besides, because a large set of HEZD parameters was fit directly to the measurement data, this approach is even more dependent on the quality of these data. Moreover, in contrast to the usually applied two-step approach, the bias of the dust-to-star flux ratio due to the assumption of a Gaussian brightness distribution within the instrumental FOV is avoided.\\
Neural networks (see \Cref{neural}) enable extremely rapid parameter estimation once trained on a representative data set and are particularly well suited for large datasets or extensive parameter studies in combination with the \(\chi_{\mathrm{red}}^{2}\) method. However, their need for careful training, and the difficulty of propagating uncertainties constrain their applicability. Thus, they are best regarded as complementary tools rather than substitutes for the one-step approach.\\
Bootstrapping (see \Cref{bootstrapping}) can be applied on top of the one-step approach as an alternative to the \(\chi_{\mathrm{red}}^{2}\) method to estimate confidence intervals via data resampling. However, a key limitation—both in this and in previous HEZD analyses—is that it does not support the simultaneous fitting of multiple free parameters. Moreover, while the one-step approach accounts for wavelength-dependent flux ratios by generating multi-wavelength brightness maps for a given set of HEZD parameters, such a functionality does not apply to bootstrapping. As a result, we are restricted to either fitting wavelength-dependent flux ratios for fixed ring parameters, or fitting the ring parameters while assuming a wavelength-independent flux ratio, indicating that a proper modeling of HEZD data requires an intrinsically wavelength-dependent approach, such as the one-step method provides.\\
 Moreover, considering again \Cref{table10}, we observe that the selected geometric model has a larger influence on the best-fit dust-to-star flux ratio than the selected fitting approaches (the results for the Gaussian models are more similar in comparison to the results for the ring models despite different fitting methods; vice-versa for the ring models). A comparison of the best-fit visibilities resulting from the different fitting approaches and geometrical models used in this study at an exemplarily wavelength of \(3.06\ \mu\mathrm{m}\) is illustrated in \Cref{Fig 12}. While the visibilities of the individual best-fit HEZD models are highly similar to each other, they all show a noticeable deviation from the visibility of the limb-darkened photosphere. Thus, despite differences in modeling and fitting approaches, all of our HEZD models yield comparable appropriate fits to the MATISSE data. Nonetheless, in line with the previous discussion, the one-step approach is preferable, particularly for better measurement data.

\section{Summary and conclusions}\label{summary}
We have presented the first observation of Fomalhaut aiming to detect the radiation of HEZD in the photometric bands \(L\)  (\(3.06 - 3.90\ \mu\mathrm{m}\)) and \(M\) (\(4.56 - 4.92\ \mu\mathrm{m}\)). The observation was carried out with MATISSE at the VLTI in medium AT configuration (four telescopes) resulting in six baselines. We applied a consistent fitting approach with respect to the underlying model geometry (one-step approach) to constrain the properties of the HEZD around Fomalhaut using the MATISSE data. Subsequently, we compared the resulting dust-to-star flux ratios at each observing wavelength  with those resulting from other fitting approaches (applying neural networks, bootstrapping, AIC, and BIC criteria) with the two-step approach  that has been usually applied in previous HEZD studies. Furthermore, we examined whether the presence of a stellar companion is compatible with the measurements. We reach the following conclusions: 
\begin{enumerate}
\item Subsets of the obtained MATISSE \(L\) band data provide a marginal detection of circumstellar radiation that could be caused by the presence of HEZD. This is confirmed by all applied fitting approaches. For the shortest observed wavelength  (\(3.06\ \mu\mathrm{m}\)), the bootstrapping fitting approach even yielded a significant (\(3\sigma\)) detection of HEZD, marking only the second detection of HEZD in the \(L\) band after \(\kappa\) Tuc (see \Cref{analysis}, \Cref{bestfit}, \Cref{bootstrapping}, and  \Cref{neural}). 
\item We derived the following best-fit HEZD parameter values for the ring model applying the one-step approach: an inner ring radius of \(0.11\ \mathrm{au}\), an outer ring radius of \(0.12\ \mathrm{au}\), a narrow dust grain size distribution around a dust grain radius constrained by \(0.53\ \mu\mathrm{m}\), a total dust mass of  \(3.25\times 10^{-10}\ \rm{M}_{\oplus}\), and an edge-on (\(90 ^\circ\)) ring inclination (see \Cref{bestfit}). 
\item A spherical shell as HEZD model instead of a ring hardly changes these best-fit parameter values (see \Cref{bestfit}).
\item The presence of a stellar companion cannot be ruled out as the explanation of the detected marginal visibility deficit and closure phases. Due to the low significance of the detection, the brightness and semi-major axis of a possible stellar companion cannot be constrained by the MATISSE data in contrast to previous studies (see \Cref{planetary}).
\item The MATISSE data and the resulting best-fit parameter values are consistent with those from previous Fomalhaut studies. While we take these previous measurements into account in our modeling, we note that this approach assumes that the observational appearance of the HEZD around Fomalhaut does not vary between observing epochs (see \Cref{bestfit} and \Cref{relation}).
\item Based on the MATISSE data and the VINCI and KIN measurements, we derived the following constraints for the model parameters: an inner ring radius in the range of \(0.10 - 0.17\ \mathrm{au}\), an outer ring radius within the range of  \(0.11 - 0.32\ \mathrm{au}\), a dust grain radius constrained by \(0.64\ \mu\mathrm{m}\), and a HEZD total mass in the range of  \(1.21\ -\ \)\(3.43\times 10^{-10}\ \rm{M}_{\oplus}\) for amorphous carbon as dust material. Thus, the origin of the circumstellar radiation around Fomalhaut is within the FOV of MATISSE (\(\sim\) 1 au). However, the MATISSE data do not yield new constraints on the model parameters (see \Cref{bestfit}).
\item In general, the dust-to-star flux ratio at each observing wavelength resulting from different fitting approaches differs and yield different parameter constraints. However, for the \mbox{MATISSE} data considered, a strong overlap of the potential dust-to-star flux ratios is obtained due to the low significance of these data: The flux ratios obtained with the two-step approach span a range from 0\% to 3.9\%. In comparison, the flux ratios derived with the one-step approach are unconstrained (i.e., no defined error limits), leading to a complete overlap with the two-step results. A similar behavior is observed for the fitting procedure based on neural networks. In contrast, the flux ratios inferred from the bootstrapping methods exhibit a narrower distribution, ranging from 0\% to 2.7\% for the Gaussian model and from 0\% to 2.9\% for the ring model (see \Cref{steps}).
\item The findings suggest that the derived dust-to-star flux ratio is more strongly influenced by the assumed geometric model than by the choice of the fitting approach (see \Cref{steps} and \Cref{comparison}).
\item A two-component structure (double-ring model) consisting of an inner hot ring and an outer warm ring is neither implied nor excluded by our analysis (see \Cref{previous}).
\end{enumerate}
Increasing the number of accurate Fomalhaut MATISSE observations simultaneously in the NIR and MIR domain leads to more significant detections of circumstellar radiation and thus also lead to tighter constraints for the HEZD properties or the attributes of a potential stellar companion. Several sequences of these combined observations also help explain a potential temporal variability in the measured flux, similar to observations of \(\kappa\) Tuc (\citealt{Ertel2016}).\\ 
Finally, the generally different dust-to-star flux ratio resulting for the different applied fitting approaches also implies different constraints on the HEZD properties around Fomalhaut, although the same measurement data were used for the fitting approaches. For example, an application of the one-step fitting approach to other HEZD systems that were previously only investigated with the two-step approach can provide different constraints on the respective HEZD properties. Moreover, the origin of the HEZD emission affects its resolvability with different interferometers. Variations in model geometry assumptions across the fitting methods impact the interpretation of spatial information in interferometric data, leading to different conclusions about its observability.

\begin{acknowledgements}
Based on observations collected at the European Organisation for Astronomical Research in the Southern Hemisphere under ESO ID 109.23HL.001 and 0109.C-0706(A). MATISSE was designed, funded and built in close collaboration with ESO, by a consortium composed of institutes in France (J.-L. Lagrange Laboratory – INSU-CNRS – Côte d’Azur Observatory – University of Côte d’Azur), Germany (MPIA, MPIfR, and University of Kiel), the Netherlands (NOVA and University of Leiden), and Austria (University of Vienna). This work is supported by the Research Unit FOR 2285 “Debris Disks in Planetary Systems” of the Deutsche Forschungsgemeinschaft (DFG). KO and SW acknowledge the DFG for financial support under contract WO 857/15-2 and AKR under 2164/15-2. FK has received funding from the European Research Council (ERC) under the European Union's Horizon 2020 research and innovation program DustOrigin (ERC-2019-StG-851622). TAS and SE are supported by the National Aeronautics and Space Administration (NASA) through Grand No. 80NSSC23K1473; TAS is also supported through NASA Grand No. 80NSSC23K0288. KT is supported by the state scholarship Schleswig-Holstein.  TDP is supported by a UKRI/EPSRC Stephen Hawking Fellowship. This research was supported in part through high-performance computing resources available at the Kiel University Computing Centre. We thank the anonymous referee for the very detailed and constructive comments that improved the quality of the paper significantly.
\end{acknowledgements}

\begin{appendix}\label{appendix}

\section{Description of bootstrapping} \label{bootstrappingmethod}

Bootstrapping (\citealt{Efron}; \citealt{Efron1982}) is a statistical technique that allows for inference from limited data sets by resampling the data multiple times. We implement a Python routine similar to the one described by \citet{Ertel} that works as follows: Given a data set with \(N\) points, we randomly sample \(M\) points with replacement, where we select \(M=N\). Subsequently, a (HEZD) model was fit \(K\) times to these \(M\) points, producing \(K\) results. Each fit is performed applying a weighted least-squared minimization using the lmfit package (\citealt{Newville}) with the inverse squared uncertainties of the individual data points as weights. For the number of samples, we used \(K=2000\), but we note that the exact number has no influence on the results as long as it is large enough.
The results are used to construct histograms for the fit flux ratio, with its median serving as the best estimate and the differences to the 16\% and 84\% quantiles providing uncertainty estimates. By resampling the data for each model fit, the inherent variability in the measurements is incorporated into the results' uncertainty. Because individual data points of VLTI measurements are usually heavily correlated, we follow \citet{Ertel} and treat all data from a single baseline as fully correlated and sample entire baselines instead of individual data points (i.e., \(N=6\)).

\section{Description of the applied neural networks}\label{Network}
Another approach to analyze the MATISSE data that we applied involved multilayer perceptrons (MLPs; see, e.g., \citealt{Krauth}; \citealt{Gallant}), which were specifically trained for this purpose. MLPs are a variant of artificial neural networks in which so-called artificial neurons are grouped into layers. Such a neural network with multiple layers is also referred to as a deep neural network. The way these layers are connected influences the architecture and functionality of the neural network. The neurons in an MLP are interconnected through weighted sums:
\begin{equation}
{a^{(l+1)}}_{i} = \sigma^{(l+1)}\Bigl(\sum^{n_l}_{j=1}{w^{(l+1)}}_{ij}{a^{(l)}}_{j}+{b^{(l+1)}}_{i}\Bigr),
\label{eq:mlp}
\end{equation}
where each of the $n_l$ neurons from the $l$-th layer is connected to every neuron in the following $(l+1)$-th layer. Here, ${a^{(l+1)}}_{i}$ is the (output) value of the $i$-th neuron in the $(l+1)$-th layer. Indices in parentheses indicate the respective layer of the variable. Additionally, $\sigma^{(l+1)}$ denotes the chosen activation function, ${w^{(l+1)}}_{ij}$ is the weight of the $i$-th neuron, which is multiplied by the $j$-th output value ${a^{(l)}}_{j}$ of the previous layer, and ${b^{(l+1)}}_{i}$ is the so-called bias of the corresponding layer. In this representation, ${a^{(1)}}_{i}$ is the input data, which may have been preprocessed, and ${a^{(\text{last layer})}}_{i}$ is the output data. The weights and biases that appear in \hyperref[eq:mlp]{Eq.~\ref*{eq:mlp}} are randomly initialized parameters that are adjusted during training of the network to optimize its performance. The activation function, applied to the sum of the bias, the weighted inputs, and the output values, introduces a non-linearity to the MLP.\\
In our model,  the HEZD is assumed to be a narrow ring seen face-on. It is defined by an inner radius $R_{\text{in}}$, a relative disk width defined by $R_{\text{out}}/R_{\text{in}}$, and an exponent for the density distribution of $\alpha=-4$. In a later step we also consider different inclinations and also different position angles of the dust distribution. It is assumed that the dust grains are composed of amorphous carbon (motivated, e.g., by \citealt{Kirchschlager2020}). We used a narrow dust grain size distribution with width $\Delta a = 0.3 a$ around the central grain size $a$.\\
Two neural networks were developed: one for the prediction of SEDs (NN1) and a second one for correlated fluxes of a face-on disk (NN2). For the optimization of the weights and biases, an Adam-Optimizer was used (\citealt{Kingma}). The chosen activation functions are the hyperbolic tangent (tanh) and the so-called Rectified Linear Unit function (ReLU), which maps positive values linearly and sets negative values to zero (see \citealt{Glorot}). For the training of each network, 50\,000 simulations were performed with DMS (\citealt{DMS}; \citealt{Stuber}) in order to derive ideal observations, another 5\,000 for validation, and 3\,000 independent simulations for testing and evaluation. In particular, single scattering and re-emission simulations were conducted, which were subsequently used to generate the corresponding SED and radial brightness profiles. To sample the data, a latin hypercube sampling method was applied (\citealt{Iman}), which provides a more even distribution of sampled random values than simple random sampling. Based on the calculated radial profiles, two-dimensional brightness maps were constructed, which were analyzed using the Python package Galario (\citealt{Tazzari}) to calculate the corresponding visibilities of the dust ring. The simulations were performed at 18 selected wavelengths between $1.5$ and $8.5\ \mu$m and at 171 linearly sampled baselengths between $30$ and $201\,$m. For the subsequent analysis of the observational data, twelve of these wavelengths and six baselines were used.\\
Both networks were trained using the python packages Tensorflow and Keras (\citealt{Chollet}; \citealt{Abadi}). The goal of a training step is to find those network parameters, which minimize a pre-defined loss function, which in this case is the mean absolute error. The final architecture of the network was then selected after a tuning procedure using the python package KerasTuner (\citealt{Malley}). The purpose of tuning is to optimize the network architecture as well as other network hyperparameters, such that the accuracy of the network is maximized on the validation set. The best hyperparameters that were found during this step, as well as further details regarding the network training, are listed in \Cref{tab:NN_hyperparameter}. The final networks were then obtained by using the best found hyperparameters. The training of the corresponding networks was additionally subdivided into two steps: initially, we used a larger learning rate and subsequently reduced it after a certain number of epochs. This allowed for a finer adjustment of the weights and biases of the network.\\
\begin{table*}[t!]
    \caption{Best found hyperparameters of the network NN1 (SEDs) and NN2 (correlated fluxes).}  
    \label{tab:NN_hyperparameter} 
    \centering
    \begin{tabular}{c c c}  
        \hline  
        Hyperparameter & NN1 for SEDs & NN2 for corr. Flux \\  
        \hline  
        Hidden layers & 4 & 6\\  
            
        \begin{tabular}[c]{@{}c@{}}Neurons in \\hidden layers\end{tabular} & \begin{tabular}[c]{@{}c@{}}360,\,350,\,\\350,\,200\end{tabular} & \begin{tabular}[c]{@{}c@{}}6440,\,1430,\,1540,\,\\3840,\,5030,\,19080\end{tabular}\\
          
        Output neurons & 18 & 18$\times$171 \\
        
        \begin{tabular}[c]{@{}c@{}}Activation\\ of hidden layers\end{tabular} & \begin{tabular}[c]{@{}c@{}}First layer: tanh;\\ then ReLU\end{tabular} & ReLU\\ 
         
        Activation of output & Linear & tanh\\
          
        Epochs & 180 & 250\\
      Learning rate & \begin{tabular}[t]{@{}c@{}}150 epochs: 2.8$\times10^{-4}$;\\ next 30 epochs: 1.8$\times10^{-7}$\end{tabular} & \begin{tabular}[t]{@{}c@{}}200 epochs: 5.1$\times10^{-5}$;\\ next 50 epochs: 1.1$\times10^{-5}$\end{tabular}
\\
        \hline
    \end{tabular}
\tablefoot{
The number of artificial neurons is arranged in ascending order of the layer numbers. The input layers of both networks have four neurons (number of model parameters). The output layer of the network NN1 has twelve neurons (number of simulated wavelengths) and the output layer of the NN2 has 18$\times$171 output neurons: one for each combination of observing wavelength and baselength.
}
\end{table*}
\begin{table*}[t!]
\caption{Considered varied parameters (first column) for the HEZD around Fomalhaut and their number of selected linearly sampled values (fourth column) within the boundaries \text{$x_{\text{min}}$} (second column) and \text{$x_{\text{max}}$} (third column).}  
    \label{tab:modell_FH}  
    \centering
    \begin{tabular}{c c c c}  
        \hline  
        Parameter & $x_{\text{min}}$ & $x_{\text{max}}$ & Grid cells \\  
          \hline
        $R_{\text{in}}$ [au] & 0.05 & 0.30 & 50\\  
        $R_{\text{out}}$ [au] & 0.051 & 0.9 & 50\\  
        $\text{Grain size}$ [$\mu$m] & 0.1 & 1.0 & 50\\  
        \hline
    \end{tabular}
\end{table*}
\end{appendix}
\end{document}